\documentclass[12pt]{article}
 
\usepackage[margin=1in]{geometry} 
\usepackage{ulem,amsmath,amsthm,amssymb}
\usepackage{graphicx}
\usepackage{enumitem}
\usepackage{setspace}
\usepackage{hyperref}

\newcommand{\E}{\mathbf{E}}
\newcommand{\Var}{\mathrm{Var}}

\usepackage{natbib}

\newcommand{\independent}{\protect\mathpalette{\protect\independenT}{\perp}}\def\independenT#1#2{\mathrel{\rlap{$#1#2$}\mkern2mu{#1#2}}}

\DeclareMathOperator*{\plim}{plim}

\usepackage[dvipsnames]{xcolor}
\usepackage{tikz}
\usepackage{indentfirst}

\newtheorem{theorem}{Theorem}
\newtheorem{corollary}{Corollary}

\newtheorem{lemma}{Lemma}
\newtheorem{assumption}{Assumption}

\title{Finitely Heterogeneous Treatment Effect in Event-study\thanks{I am deeply grateful to Stéphane Bonhomme, Christian Hansen and Azeem Shaikh, who have provided me invaluable support and insight. I would also like to thank Manasi Deshpande, Ali Hortaçsu, Guillaume Pouliot, Max Tabord-Meehan, Alex Torgovitsky, Martin Weidner and the participants of the metrics advising group at the University of Chicago and the Nuffield Econometrics Seminar at the University of Oxford for their excellent feedback. I acknowledge the support from the European Research Council through the grant ERC-2018-CoG-819086-PANEDA. Any and all errors are my own. }
}
\author{Myungkou Shin\thanks{School of Social Sciences, University of Surrey. email: \href{mailto:m.shin@surrey.ac.uk}{m.shin@surrey.ac.uk}}
}
\date{September 26, 2024}

\begin{document}

\maketitle

\begin{abstract}
A key assumption of the differences-in-differences designs is that the average evolution of untreated potential outcomes is the same across different treatment cohorts: a parallel trends assumption. In this paper, we relax the parallel trend assumption by assuming a latent type variable and developing a \textit{type-specific} parallel trend assumption. With a finite support assumption on the latent type variable and long pretreatment time periods, we show that an extremum classifier consistently estimates the type assignment. Based on the classification result, we propose a type-specific diff-in-diff estimator for type-specific ATT. By estimating the type-specific ATT, we study heterogeneity in treatment effect, in addition to heterogeneity in baseline outcomes. 
\end{abstract}

\noindent \hspace{8mm} \textbf{Keywords}: event-study, difference-in-differences, panel data, heterogeneity, \\
\text{ } \hspace{28.5mm} classification, $K$-means clustering

\noindent \hspace{8mm} \textbf{JEL classification codes}: C13, C14, C23

\vspace{10mm} 

\pagebreak 

\section{Introduction}

The event-study design, which utilizes the variation in the timing of an event, is an empirical framework whose popularity among empirical researchers has risen tremendously over the time. In applied microeconomics, the event-study design is most often implemented with the difference-in-differences (diff-in-diff) approach. A key identifying assumption of the diff-in-diff style event-study research design is the parallel trend assumption: temporal differences of untreated potential outcomes are mean independent of treatment timing. 

The traditional parallel trend assumption that assumes parallel trend with unconditional means is limited in its scope; it is directly violated when there exist heterogeneous patterns of time trends across units. A widely used alternative to this unconditional parallel trend assumption is to assume a conditional parallel trend with observable conditioning covariates (see \citet{Abadie2005,CS} among others). Following the same spirit, we also assume conditional parallel trend assumption to relax the unconditional parallel trend assumption, but assume that the conditioning variable is latent. Throughout the paper, we call the conditioning variable a latent `type' variable and call the conditional parallel trend a `type-specific parallel trend' assumption. To illustrate the assumption, let us consider a simple two time periods case: with a latent type variable $k_i$, \begin{align}
    \E[ Y_{i2}(\infty) - Y_{i1}(\infty) | k_i, D_{i}=1] = \E[ Y_{i2}(\infty) - Y_{i1}(\infty) | k_i, D_{i}=0]. \label{eq:type-specificPT}
\end{align}
$Y_{it}(\infty)$ denotes untreated potential outcome for unit $i$ at time $t$ and $D_{i}$ denotes an indicator whether unit $i$ is treated at time $t=2$. When $k_i$ is observable, Equation \eqref{eq:type-specificPT} is the conditional parallel trend assumption as in \citet{Abadie2005,CS}.

In this paper, allowing for the conditioning variable $k_i$ to be latent comes at three additional requirements on the model and data. Firstly, we assume that the latent type variable $k_i$ has a finite support. Secondly, we assume that the types are sufficiently separated in terms of pretreatment outcomes. Lastly, we assume that the number of pretreatment periods grows to infinity. In this sense, the type-specific parallel trend assumption can be understood as an alternative to the unconditional parallel trend or the conditional parallel trend with observable conditioning covariates; it allows for heterogeneity patterns that may not be captured by the observable information, at the price of assuming that the heterogeneity is finite and the pretreatment time series is long. Moreover, the finite support assumption motivates the use of a simple within-type diff-in-diff style estimator, which is potentially less susceptible to the overlap problem compared to the conditional diff-in-diff estimators with a continuous conditioning variable.

The finite support assumption gives our framework a unique merit; it helps us analyze the patterns of treatment effect heterogeneity. While the type-specific parallel trend framework of this paper does not put any restriction on the treatment effect heterogeneity, thus allowing for fully flexible treatment effect heterogeneity, the finite support assumption provides a model-based stratifying structure that can be helpful in summarizing the treatment effect heterogeneity. Let $Y_{i2}(2)$ denote the treated potential outcome of unit $i$ at time $t=2$. The existing literature mostly focuses on estimating the conditional expectation of $Y_{i2}(2) - Y_{i2}(\infty)$ given some observable information: e.g., $\E \left[ Y_{i2}(2) - Y_{i2}(\infty) | X_i, D_i=1 \right]$. With the type-specific parallel trend framework, we document treatment effect heterogeneity along the latent type variable $k_i$: $$
\E \left[ Y_{i2}(2) - Y_{i2}(\infty) | k_i, D_i = 1 \right].\footnotemark
$$ 
The stratifying structure induced by the finite type can be particularly helpful for policy recommendation when policymakers have little to zero constraints in policy assignment but do not have perfect information for individual-level responsiveness; it provides a feasible alternative to individual-level treatment assignment, by identifying the treatment effect at the type level. \footnotetext{The treatment effect parameter is comparable to those defined in \citet{Abadie2005,SZ,CS} when $k_i$ is observable.}

To apply the type-specific parallel trend framework to datasets, we propose a two-step estimation procedure. In the first step, we classify units into the finite number of types, using the $K$-means clustering algorithm on the long pretreatment periods. Given the classification result, the second step of the estimation procedure estimates the treatment effect, using the estimated types as given.\footnotemark \footnotetext{This two-step property of the estimation procedure closely relates to the stratification exercise used in estimating subpopulation treatment effect (see \citet{ACW}). The goal of the stratification (i.e. classification in this paper's terminology) is to find groups of units whose (estimated) counterfactual untreated outcomes are similar. The type-specific parallel trend assumption directly relates to this since under the type-specific parallel trend assumption, units with the same type share the same time trend.} In the estimation step, a variety of existing estimation strategies can be used by treating the type as a given categorical variable: \citet{CD,BJS,CS,SA}. As with \citet{CD} and \citet{CS}, our `type-specific diff-in-diff' estimator takes an average of the canonical diff-in-diff estimates with two time periods and two treatment timings. 

To discuss asymptotic properties of the treatment effect estimators, we first show that the probability of first-step misclassification goes to zero when the number of pretreatment time periods grows at a polynomial rate of the number of units. Given that the number of pretreatment time periods grows sufficiently fast compared to the number of units, the type-specific diff-in-diff estimators are consistent and asymptotically normal under some regularity conditions. These asymptotic results are supported by Monte Carlo simulations. 


This paper contributes to the large literature of panel data models where interactive fixed-effects models are used to control for unit heterogeneity across treatment cohorts, relaxing the two-way fixed-effect (TWFE) specification used in the parallel trend approach: see \citet{ADH10,Aetal,ABDIK,HCW,FHS,X,CHLZ,CKarami,JS} among others. The interactive fixed-effect model often assumes a factor model for the interactive fixed-effects; our framework can be thought of as a special case of the factor model with a finite support on the factor. \citet{JS} is closest to this paper in that it takes the same approach; however, \citet{JS} neither explores treatment effect heterogeneity nor develop a full asymptotic theory. 
As with this paper, most of the papers in the interactive fixed-effect model literature rely on large pretreatment periods. Notable exceptions are \citet{CKarami} and \citet{FHS}. 
In both papers, informative external variables or instruments are used, instead of assuming long pretreatment periods. 

This paper also closely relates to the rapidly growing literature on heterogeneous treatment effect: see \citet{CD,SA,CS,G,BJS,BLW,GHK}, among others. \citet{CS} is particularly close to this paper in the sense that they also consider a conditional parallel trend assumption. This literature discusses the negative weighting problem that arises in the standard TWFE specification when there is treatment effect heterogeneity across units. We build upon this literature and construct the type-specific diff-in-diff estimator to be robust to the treatment effect heterogeneity. 

The rest of the paper is organized as follows. In Section \hyperlink{Sec2}{2}, we formally discuss the type-specific parallel trend assumption. In Section \hyperlink{Sec3}{3}, we propose the two-step estimation procedure for treatment effect estimation. In Section \hyperlink{Sec4}{4}-\hyperlink{Sec5}{5}, we discuss the asymptotic properties of the estimator. In Section \hyperlink{Sec6}{6}, we present Monte Carlo simulation results on the finite-sample performance of the estimator. In Section \hyperlink{Sec7}{7}, we provide an empirical illustration of the type-specific diff-in-diff estimator by revisiting \citet{L}. 

\section{Model}

\hypertarget{Sec2}{}

In the main model of the paper, an econometrician observes a panel data with a binary treatment: $\left\lbrace \left\lbrace Y_{it}, D_{it} \right\rbrace_{t=-T_0-1}^{T_1-1}\right\rbrace_{i=1}^n$. $Y_{it}$ is the outcome variable for unit $i$ at time $t$ and $D_{it} \in \{0,1\}$ is the binary treatment variable for unit $i$ at time $t$. $D_{it}$ follows the staggered adoption scheme; $D_{it} \leq D_{it+1}$. $E_i = \min \{t: D_{it}=1\}$ denotes the treatment timing of unit $i$; this is the same notation as in \citet{SA} and equivalent with treatment `group' notation $G$ in \citet{CS}. For never-treated units, let $E_i = \infty$. There are $n$ units and $T+1=T_0+T_1+1$ time periods, with the unit index ranging $i=1, \cdots, n$ and the time index ranging $t=-T_0-1,\cdots,T_1-1$. $T_0+1$ denotes the number of population pretreatment periods and $T_1$ denotes the number of population treatment periods; $\sum_{i=1}^n D_{it} = 0$ for all $t < 0$. $t < 0$ denotes pretreatment periods at the population level and $t \geq 0$ denotes treatment periods at the population level. $T_1$ is fixed. Throughout the paper, we use the potential outcome framework to discuss treatment effect: \begin{align*}
    Y_{it} &= Y_{it}(E_i). 
\end{align*}
$Y_{it}(e)$ is the potential outcome of unit $i$ at time $t$ when their treatment timing is $e$. Thus, for some $Y_{it}(e)$, $t<e$ means untreated potential outcome and $t \geq e$ means treated potential outcome. 

The key assumption of this paper is that there exists a unit-level latent type variable. Conditional upon the latent type, the parallel trend assumption and the no anticipation assumption hold.  
\begin{assumption}
\hypertarget{A1}{} \renewcommand{\baselinestretch}{0.7}
\textsc{(Type-Specific Parallel Trend)} There exists a latent type variable $k_i$ such that for any $t,s$ \begin{align*} 
     \E \left[ Y_{it}(\infty) - Y_{is}(\infty) | k_i, E_i \right] = \E \left[ Y_{it}(\infty) - Y_{is}(\infty) | k_i \right]
\end{align*}
\end{assumption}

\begin{assumption}
\hypertarget{A2}{} \renewcommand{\baselinestretch}{0.7}
\textsc{(No Anticipation)} for any $t < e$ \begin{align*} 
     \E \left[ Y_{it}(e) - Y_{it}(\infty) | k_i, E_i \right] = 0
\end{align*}
\end{assumption}

\noindent Assumption \hyperlink{A1}{1} assumes that there is some latent type variable $k_i$, conditioning on which the time trend of the never-treated potential outcome $Y_{it}(\infty) - Y_{is}(\infty)$ is mean independent of the treatment timing $E_i$.\footnotemark \footnotetext{The parallel trend is assumed for every time period, across every treatment timing; when $K=1$, this is equivalent to the conventional parallel trend assumption assumed in \citet{CD,SA}, but different from \citet{CS}, which restricts the window of time periods to apply the parallel trend to.} Assumption \hyperlink{A2}{2} assumes that the never-treated potential outcome $Y_{it}(\infty)$ and the pretreatment potential outcome $Y_{it}(e)$ for $t < e$ have the same conditional mean, given the latent type $k_i$ and the treatment timing $E_i$.

In addition to Assumptions \hyperlink{A1}{1}-\hyperlink{A2}{2}, we assume the following two assumptions, to have a type classification results on $k_i$.  \begin{assumption}
\hypertarget{finite}{} \textsc{(Finite Support)} \renewcommand{\baselinestretch}{0.7} \begin{align*}
    k_i \in \{1, \cdots, K\}. 
\end{align*}
\end{assumption} 

\begin{assumption}
\hypertarget{separation}{} \textsc{(Well-Separated Types)} \renewcommand{\baselinestretch}{0.7} whenever $k \neq k'$, \begin{align*}
    \frac{1}{T_0} \sum_{t=-T_0}^{-1} \Big( \E \left[ Y_{it} (\infty) - Y_{it-1} (\infty) | k_i = k \right] - \E \left[ Y_{it} (\infty) - Y_{it-1} (\infty) | k_i = k' \right] \Big)^2 \to c(k,k') > 0 
\end{align*}
as $T_0 \to \infty$. 
\end{assumption} 

\noindent Assumption \hyperlink{finite}{3} assumes that the latent type variable $k_i$ has a finite support, as in the group fixed-effect literature or the finite mixture literature. Assumption \hyperlink{separation}{4} assumes that the types are well-separated in terms of $\E[Y_{it}(\infty) - Y_{it-1}(\infty) | k_i=k]$; for any two different types; the $l_2$ norm of the difference between their conditional means is strictly nonzero. Note that the separation assumption is in relation to time trends of the never-treated potential outcomes $Y_{it}(\infty)$. From Assumptions \hyperlink{A1}{1}-\hyperlink{A2}{2}, we have \begin{align}
\E \left[ Y_{it}(\infty) - Y_{it-1}(\infty) | k_i =k \right] = \E \left[ Y_{it}(e) - Y_{it-1}(e) | k_k=k, E_i=e \right] \label{eq:ATTid}
\end{align}
whenever $t < e$. Thus, Assumption \hyperlink{separation}{4} can be extended to the pretreatment outcomes of all units regardless of their treatment timings.

Given the finite type structure, let us define type-specific treatment effect parameters. Fix two time periods $(s,t)$ and a treatment cohort $\{i:E_i=e\}$ such that $s < e \leq t$. The type-cohort-and-time specific average treatment effect on treated units (ATT) for time $t$, type $k$ and treatment timing $e$ can be written as follows: \begin{align}
ATT_t(k,e)&=\E \left[ Y_{it}(e) - Y_{it}(\infty) | k_i=k, E_i=e \right] \label{eq:fullCATT} \\
&= \E \left[ Y_{it}(e) - Y_{is}(\infty) | k_i=k, E_i=e \right] - \E \left[ Y_{it}(\infty) - Y_{is}(\infty) | k_i=k, E_i=e \right] \notag \\
&= \E \left[ Y_{it} - Y_{is} | k_i=k, E_i=e \right] - \E \left[ Y_{it} - Y_{is} | k_i=k, E_i > t \right]. \notag 
\end{align}
The third equality holds from Assumptions \hyperlink{A1}{1}-\hyperlink{A2}{2} and thus $ATT_t(k,e)$ is written as a function of the conditional moments of $\{Y_{it}\}_{t}$ given the latent type variable $k_i$ and the treatment timing variable $E_i$. $ATT_t(k,e)$ is a type-specific version of the treatment effect estimands used in the diff-in-diff literature: see \citet{CS,SA} for more. 

The type-cohort-and-time specific ATT parameter in \eqref{eq:fullCATT} takes treatment timing $E_i$ as a conditioning variable and focuses on a specific time period $t$. The full-fledgedness of the type-cohort-and-time specific ATT is useful when the researcher is interested in treatment effect heterogeneity across both time periods and types. Though both dimensions of the treatment effect heterogeneity may be of interest depending on contexts, we focus on an aggregated ATT parameter in this paper, to highlight the treatment effect heterogeneity across types. To aggregate, we take the average of \eqref{eq:fullCATT} across $(t,e)$ while maintaining the relative treatment timing $t-e$ fixed: for some $r \geq 0$, \begin{align}
    \beta_r(k) &:= \sum_{e=0}^{T-1-r} \frac{\Pr \left\lbrace E_i = e |k_i=k \right\rbrace}{\Pr \left\lbrace E_i \leq T_1 - r |k_i=k\right\rbrace} \cdot \E \left[ Y_{i,e+r}(e) - Y_{i,e+r}(\infty) | k_i=k, E_i=e \right]. \label{eq:dCATT}
\end{align}
$\beta_r(k)$ is the $r$-times lagged ATT for type $k$. In aggregating $ATT_t(k,e)$, we use the weights discussed in \citet{CS,SA}. We call $\beta_r(k)$ as `dynamic'  type-specific ATT since the parameter compares instantaneous treatment effect and long-run treatment effect for each type $k$, as we compare $\beta_r(k)$ with $r \approx 0$ and $\beta_r(k)$ with $r \gg 0$. 

It is straightforward to see that $ATT_t(k,e)$ and $\beta_r(k)$ are identified from the distribution of $\left( \left\lbrace Y_{it} \right\rbrace_{t=-T_0-1}^{T_1-1}, E_i, k_i \right)$ whenever their respective overlap conditions discussed in Section \hyperlink{Aol}{C.1} are satisfied and $\{k_i\}_{i=1}^n$ is observed for a given $T_0$. Though the type assignment $\{k_i\}_{i=1}^n$ is not directly observed from data, the following lemma shows that the treatment effect parameters are identified when $T_0 \to \infty$ and the pretreatment time series shows weak dependence.

\begin{lemma}
    \hypertarget{L1}{} Assumptions \hyperlink{A1}{1}-\hyperlink{separation}{4} hold. Also, $\Pr \left\lbrace k_i = k \right\rbrace > 0$ and $$
    \lim_{T_0 \to \infty} \frac{1}{T_0} \sum_{t=-T_0}^{-1} \big( \E \left[ Y_{it}(\infty) - Y_{it-1}(\infty) | k_i=k \right] \big)^2
    $$
    exists for each $k$ and $$
    \plim_{T_0 \to \infty} \frac{1}{T_0} \sum_{t=-T_0}^{-1} a_t \big( Y_{it}(E_i) - Y_{it-1}(E_i) - \E \left[ Y_{it}(\infty) - Y_{it-1}(\infty) |k_i \right] \big) = 0
    $$
    for any $\{a_t\}_{t=-T_0}^{-1}$ such that $\lim_{T_0 \to \infty} \frac{1}{T_0} \sum_{t=-T_0}^{-1} {a_t}^2$ is finite. Then, $ATT_t(k,e)$ and $\beta_r(k)$ are identified from the distribution of $\left( \left\lbrace Y_{it}\right\rbrace_{t=-\infty}^{T_1-1}, E_i \right)$ when their respective overlap conditions are satisfied. 
\end{lemma}

\noindent Specifics of the overlap condition are discussed in Section \hyperlink{Aol}{C.1} of the Appendix. The proof for Lemma \hyperlink{L1}{1} is found in Section B of the Supplementary Appendix. Lemma \hyperlink{L1}{1} is an identification counterpart of the type classification consistency result in Section \hyperlink{Sec4}{4}: Theorem \hyperlink{T1}{1}. 

\section{Estimation}

\hypertarget{Sec3}{}

The estimation procedure is two-step. The first step is to estimate the type using the $K$-mean minimization problem. The second step is to take the estimated type as given and estimate ATT. To describe the estimation procedure, let us adopt the following notations: \begin{align*}
    \gamma &:= \left( k_1, \cdots, k_n \right)\in \Gamma, \\
    \Gamma &:= \left\lbrace 1, \cdots, K \right\rbrace^n, \\
    \delta &:= \left\lbrace \delta_t(k) \right\rbrace_{t,k} 
\end{align*}
$\gamma$ is a $n \times 1$ vector of a type assignment while $k_i$ denotes the type of unit $i$. $\Gamma$ is a set of all possible type assignments where $n$ units are assigned to $K$ different types. $\delta$ is a collection of $\delta_t(k)$, the type-specific time trend given time $t$ and type $k$: $$
\delta_t(k) = \E \left[ Y_{it}(\infty) - Y_{it-1}(\infty) | k_i=k \right].
$$ 

In the classification step, we only use a subset of the given data: population pretreatment periods. With the population pretreatment periods, we construct an objective function:
\begin{align}
    \widehat{Q} ({\delta}, \gamma) &= \frac{1}{n T_0} \sum_{i=1}^n \sum_{t=-T_0}^{-1}  \left( {Y}_{it} - Y_{it-1} - {\delta}_t (k_i) \right)^2 \label{eq:Kmeans} \\
    \intertext{and the resulting first-step classifier is }
    \left( \hat{\delta}, \hat{\gamma} \right) &= \arg {\min}_{\left( \delta, \gamma \right) \in \mathcal{D} \times \Gamma} \widehat{Q} \left( {\delta}, \gamma \right). \label{eq:estimator}
\end{align}
$\mathcal{D}=[-M,M]^{T_0 K}$ with some $M>0$. The minimization problem in \eqref{eq:Kmeans} is called $K$-mean minimization problem; the solution to the $K$-means minimization problem is a grouping structure with $K$ groups, defined with $K$ centeroids. In our minimization problem \eqref{eq:Kmeans}, the centeroids are denoted with $\left\lbrace \delta_t(1) \right\rbrace_{t< 0}, \cdots, \left\lbrace \delta_t(K) \right\rbrace_{t < 0}$ and the grouping structure is denoted with $k_1, \cdots, k_n$. 

The algorithm that we use to obtain \eqref{eq:estimator} is a conventional $K$-means clustering algorithm. Given an initial type assignment ${\gamma}^{(0)} = \left( k_1^{(0)}, \cdots, k_n^{(0)} \right)$,
\begin{enumerate}
    \item \textbf{(update $\delta$)} Given the type assignment $\gamma^{(s)}$ from the $s$-th iteration, estimate $\hat{\delta}_t^{(s)}(k)$ by letting $$
    \hat{\delta}_t^{(s)}(k) = \frac{\sum_{i=1}^{n} \left( Y_{it} - Y_{it-1} \right) \mathbf{1}\{k_i^{(s)}=k\}}{\sum_{i=1}^{n} \mathbf{1}\{k_i^{(s)}=k\}}
    $$
    whenever the denominator is not zero. 
    
    \item \textbf{(update $\gamma$)} Update $k_i^{(s)}$ for each $i$ by letting $k_i^{(s+1)}$ be the solution to the following minimization problem: for $i = 1, \cdots, N$, $$
    \min_{k\in \{1,\cdots,K\}} \sum_{t=-T_0}^{-1} \left( {Y}_{it} - Y_{it-1} - \hat{\delta}_t^{(s)} (k)  \right)^2. 
    $$
    
    \item Repeat Step 1-2 until Step 2 does not update $\hat{\gamma}$, or some stopping criterion is met. For stopping criterion, one can set a maximum number of iteration or a minimum update in $\hat{\delta}^{(s)}$: set $S$ and $\varepsilon$ such that the iteration stops when $$
    s \geq S \hspace{3mm} \text{or} \hspace{3mm} \left\| \hat{\delta}^{(s)} - \hat{\delta}^{(s-1)} \right\|_\infty \leq \varepsilon.    
    $$
\end{enumerate}

The iterative algorithm proposed here has two stages. In the first stage, the algorithm estimates $\delta$ by taking sample means. In the second stage, the algorithm reassigns a type for each unit, by finding the type that minimizes the distance between $\{Y_{it}-Y_{it-1}\}_{t<0}$ and $\{\delta_t(k)\}_{t<0}$. The algorithm quickly attains a local minimum of the minimization problem \eqref{eq:Kmeans}. In the application we used in Section \hyperlink{Sec7}{7}, the algorithm mostly converged within 20 iterations. 

Since the iterative algorithm does not conduct an exhaustive search, it may not converge to a global minimum; the computational burden of the exhaustive search is extremely heavy since the space for the type assignment has cardinality of $n^K$. Thus, we recommend that a random initial type assignment be drawn multiple times and the associated local minima be compared. 

Another concern is the choice of $K$. So far, the number of types $K$ has been treated as known. When there is no natural choice for $K$, an information criterion can be used to estimate the number of type $K$ (see \citet{BN,BM,JS}). To construct the information criterion, first we set $K_{\max}$, an upper bound on the number of types. Then, we estimate the error variance using the classification with $K_{\max}$: with $\Gamma(K) = \{1,\cdots, K\}^n$, $$
\hat{\sigma}^2 = \min_{\delta, \gamma \in \Gamma(K_{\max})} \widehat{Q} (\delta, \gamma)
$$
Given the estimate, we estimate $K$ to be the minimizer of the Bayesian information criterion: $$
\hat{K} = \arg \min_{K \leq K_{\max}}  \left( \min_{\delta, \gamma \in \Gamma(K)} \widehat{Q}(\delta, \gamma) + \hat{\sigma}^2 \frac{K T_0 + n}{nT_0} \log n T_0 \right).
$$
See the Supplementary Appendix for more discussion. 

Given the first-step result, the type-specific diff-in-diff estimator for $ATT_t(k,e)$ defined in \eqref{eq:fullCATT} can be constructed by taking sample means for each type: $$
\widehat{ATT}_t(k,e) = \sum_{i=1}^n \left( Y_{it} -Y_{i,e-1} \right) \left( \frac{ \mathbf{1}\{\hat{k}_i=k,E_i=e\}}{\sum_{i=1}^n \mathbf{1}\{\hat{k}_i=k,E_i=e\}} - \frac{\mathbf{1}\{\hat{k}_i=k,E_i>t\}}{\sum_{i=1}^n \mathbf{1}\{\hat{k}_i=k,E_i>t\}} \right).\footnotemark
$$
In the case of the dynamic ATT parameter $\beta_r(k)$ defined in \eqref{eq:dCATT}, the estimator is \begin{align}
\hat{\beta}_r(k) &= \sum_{e \leq T_1-1-r} \frac{\hat{\mu}(k,e)}{\sum_{e' \leq T_1-1-r} \hat{\mu}(k,e')} \cdot \widehat{ATT}_{e+r}(k,e)  \label{eq:dCATTest}
\end{align}
where $\hat{\mu}(k,e) = \frac{1}{n} \sum_{i=1}^n \mathbf{1}\{\hat{k}_i=k, E_i=e\}$. $\hat{\mu}(k,e)$ is the estimator for the probability $\mu(k,e)= \Pr \left\lbrace k_i=k, E_i=e \right\rbrace$. 

\footnotetext{As discussed in \citet{CS}, there is no straightforward choice in which time differences to use in a diff-in-diff type approach. Though the estimator described above takes one period before the treatment timing to construct a time difference, other choices are equally valid as long as the parallel trend assumption holds for every time period. \citet{RSjpem} discusses efficiency of these diff-in-diff type estimates when the treatment timing is truly random. More discussion on this is given the \hyperlink{Asec1}{Appendix}. Also, the estimator uses the not-yet-treated units as control units; alternatively, one can use never-treated units or any other subset of $\{i: E_i > t\}$.} 

Similarly, we can extend $\hat{\beta}_r(k)$ for $r < -1$ and construct estimators for $$
\beta_r(k) = \sum_{e=0}^{T-1} \frac{\Pr \left\lbrace E_i = e |k_i=k\right\rbrace}{\Pr \left\lbrace E_i \leq T_1 - r |k_i=k\right\rbrace} \cdot \E \left[ Y_{i,e+r}(e) - Y_{i,e+r}(\infty) | k_i=k, E_i=e \right]
$$
for some $r < -1$. From Assumptions \hyperlink{A1}{1}-\hyperlink{A2}{2}, $\beta_r(k)= 0$ whenever $r < -1$. Thus, we can use $\hat{\beta}_r(k)$ for $r < -1$ to test Assumptions \hyperlink{A1}{1}-\hyperlink{A2}{2}, equivalent to the widely used `no pretreatment test' in the event-study literature. 

\section{Asymptotic Theory}

\hypertarget{Sec4}{}

In this section, we discuss the asymptotic properties of the estimator proposed in Section \hyperlink{Sec3}{3}. As we extend the main model and the asymptotic results of this section in Section \hyperlink{Sec5}{5} and Sections \hyperlink{At}{C.2-4} of the Appendix, the proofs are implied by the more general results. The proofs for the general results and how they connect to the asymptotic results of this section are discussed in Sections C-D of the Supplementary Appendix.  

Firstly, to derive the classification result for the type estimator defined in \eqref{eq:estimator}, let us adopt following assumptions. 

\begin{assumption}
\hypertarget{A5}{} 
With some $M>0$,
\begin{itemize}
    \item[\textbf{a.}] (iid across units) $\left( \left\lbrace Y_{it}(e) \right\rbrace_{e,t}, E_i, k_i \right) \overset{\text{iid}}{\sim} F$. 
    
    \item[\textbf{b.}] (finite moments) For every $e$, $t$ and $k$, $\E \left[ Y_{it}(e)^4 \big| k_i=k \right] \leq M$.
    
    \item[\textbf{c.}] (long pretreatment) $T_0 \to \infty$ as $n \to \infty$. 

    \item[\textbf{d.}] (no measure zero types) For all $k \in \{1, \cdots, K\}$, $\Pr \left\lbrace k_i = k \right\rbrace > 0$
    
    \item[\textbf{e.}] (weakly dependent, thin-tailed errors) With some positive constant $d_1$ and $a$, $$
    \Big\lbrace Y_{it}(e) - Y_{it-1} (e)- \E \left[ Y_{it} (\infty) - Y_{it-1}(\infty) | k_i \right] \Big\rbrace_{t=-T_0}^{-1}
    $$
    is strongly mixing with mixing coefficient $\alpha[t]$ such that $\alpha[t] \leq \exp ( - a t^{d_1} )$ uniformly over $e$. Also, with some positive constant $d_2$ and $b$, $Y_{it}(e)$ satisfies the following tail probability: for any $y > 0$,
    $$
    \Pr \left\lbrace \left| Y_{it}(e) - \E \left[ Y_{it} (\infty) | k_i \right] \right| \geq y \right\rbrace \leq \exp \left( 1 - \left( y/b \right)^{d_2} \right)
    $$
    uniformly over $e$ and $t <0$.
\end{itemize}
\end{assumption}

\noindent Assumption \hyperlink{A5}{5-c} assumes that the number of population pretreatment periods $T_0$ grows to infinity as $n$ goes to infinity. Assumption \hyperlink{A5}{5-d} assumes that each type realizes with positive probability. Assumption \hyperlink{A5}{5-e} assumes that for $t < 0$, tail probability of $Y_{it}(e) - \E [ Y_{it}(\infty)|k_i]$ goes to zero exponentially and the first difference of $Y_{it}(e) - \E [ Y_{it} (\infty)|k_i]$ is weakly dependent in the sense that it is strongly mixing with mixing coefficient decreasing exponentially in $t$. 

\begin{theorem}
\hypertarget{T1}{} Assumptions \hyperlink{A1}{1}-\hyperlink{A5}{5} hold. Then, up to some permutation on $\{1,\cdots, K\}$, \begin{align*}
    \Pr \left\lbrace \sup_i \mathbf{1}{\{\hat{k}_i \neq k_i^0\}} > 0 \right\rbrace &= o (n{T_0}^{-\nu}) + o(1) \hspace{3mm} \forall \nu >0
\end{align*} 
as $n \to \infty$. 
\end{theorem}

\noindent Theorem \hyperlink{T1}{1} puts a bound on the misclassification probability; the rate is identical to the rate found in the group fixed-effect literature. 

The classification of $n$ units into $K$ types is a crucial part of the estimation procedure that the performance of the treatment effect estimators depends on. Consider a very simple case where $K=2$ and model the untreated potential outcomes as follows: for $t \leq 0$, $$
Y_{it}(\infty) = \delta(k_i) + U_{it}, \hspace{8mm} U_{it} \overset{iid}{\sim} \mathcal{N} (0,1). 
$$
WLOG let $\delta(1) < \delta(2)$. Find that $\bar{Y}_i(\infty) = \frac{1}{T_0+1} \sum_{t=-T_0-1}^{-1} Y_{it}(\infty) \sim \mathcal{N} \left( \delta(k_i), \frac{1}{T_0+1} \right)$. It is easy to see that for any fixed $T_0$, \begin{align*}
    \Pr \left\lbrace \bar{Y}_i(\infty) \geq \bar{Y}_j(\infty) | k_i = 1, k_j = 2\right\rbrace &= \Pr \left\lbrace \bar{U}_i - \bar{U}_j \geq \delta(2) - \delta(1) | k_i = 1, k_j = 2\right\rbrace \\
    &= \Phi \left( \sqrt{\frac{T_0+1}{2}} \Big( \delta(2) - \delta(1) \Big) \right)
\end{align*} 
is nonzero, with $\Phi$ being the distribution function of the standard normal $\mathcal{N}(0,1)$; the probability of imperfect classification is nonzero. Thus, we need large pretreatment periods $(\Leftrightarrow T_0 \gg 0)$, in addition to the strong separation $(\Leftrightarrow \delta(2) -\delta(1) > 0 )$.\footnotemark \footnotetext{By evaluating the CDF function $\Phi$, we can see that ${T_0}^{\nu} \Phi \left( \sqrt{\frac{T_0+1}{2}}\left( \delta(2) - \delta(1) \right) \right)$ goes to zero for any $\nu > 0$ as $T_0$ grows, as stated in Theorem \hyperlink{T1}{1}.} When we do not have both conditions satisfied and thus units are potentially misclassified, the treatment effect estimator suffers from a non-classical measurement error problem.\footnotemark \footnotetext{The measurement error problem may be severe for a type-specific ATT even when the misclassification rate is small and the tail of the potential outcome distribution is thin, if the within-type overlap is weak. On the other hand, when the misclassification rate is negligible compared to the within-type shares of treated units and control units, the bias will be small.}

Given the long pretreatment, the bound on the misclassification probability from Theorem \hyperlink{T1}{1} can be used to derive asymptotic properties of the type-specific diff-in-diff estimator from \eqref{eq:dCATTest}. 

\begin{corollary}

\hypertarget{C1}{}

Assumptions \hyperlink{A1}{1}-\hyperlink{A5}{5} hold. For any $t,s$ and $e$, $\Var \left( Y_{it}(e) - Y_{is} (e) | k_i, E_i \right) > 0$. There exists some $\nu^*>0$ such that $n / {T_0}^{\nu^*} \to 0$ as $n \to \infty$. Then, for any $k$ and $r$ with some permutation on $\{1,\cdots, K\}$, $$
\sqrt{n} \left( \hat{\beta}_r(k) - \beta_r(k) \right) \xrightarrow{d} \mathcal{N} \left( 0, {\sigma_{\beta_r(k)}}^2 \right)
$$    
with some ${\sigma_{\beta_r(k)}}^2 >0$, as $n \to \infty$, when $k$ and $r$ satisfy the corresponding overlap condition.
\end{corollary}

\noindent \textit{Remark 1.} The asymptotic variance has a consistent estimator, whose expression is given in the Supplementary Appendix, along with the proof of Corollary \hyperlink{C3}{3}. 

\vspace{3mm}

\noindent \textit{Remark 2.} In formulating the dynamic ATT parameter $\beta_r(k)$, treatment timing distribution is used as weights. Similar asymptotic results as in Corollary \hyperlink{C1}{1} hold for many other choices of weights: e.g. uniform weights across treatment timing. 

\section{Extension to the Model with Covariates}

\hypertarget{Sec5}{}

\subsection[Introducing control covariates]{Introducing control covariates $X_{it}$}

In this section, we extend our main model by adding observed covariates $X_{it} \in \mathbb{R}^p$ to the model. The control covariates $X_{it}$ gives us an extra source of heterogeneity in outcomes across different units and different times. For the classification to be successful, we need to decompose the variation in the outcome variable into the variation from the control covariates $X_{it}$ and the variation from the latent type $k_i$. For that end, we assume the following linear model for untreated potential outcome: for $t < 0$,  \begin{align}
    Y_{it}(E_i) - Y_{it-1}(E_i) &= \delta_t(k_i) + {X_{it}}^\intercal \theta + U_{it}. \label{eq:ParModel} 
\end{align}
\noindent Note that the interpretation of $\delta_t(k)$ is changed. In the linear model \eqref{eq:ParModel}, $\delta_t(k)$ is not the conditional mean of first-differenced potential outcome anymore since there exists ${X_{it}}^\intercal \theta$. Thus, we call $\delta_t(k)$ the type-specific time fixed-effects. The type-specific time fixed-effects explains heterogeneity across units that is not explained by the (linear) observable control covariates $X_{it}$. 

Given the model \eqref{eq:ParModel}, we can construct a similar objective function from before and solve the $K$-means minimization problem for classification: $$
\left( \theta, \delta, \gamma \right) = \arg \min_{\theta, \delta, \gamma} \frac{1}{nT_0} \sum_{i=1}^n \sum_{t=-T_0}^{-1} \Big( Y_{it} - Y_{it-1} - \delta_t(k_i) - {X_{it}}^\intercal \theta \Big)^2.
$$
The objective function includes $X_{it}$. Given an initial type assignment ${\gamma}^{(0)} = \left( k_1^{(0)}, \cdots, k_N^{(0)} \right)$,
\begin{enumerate}
    \item \textbf{(update $\theta$ and $\delta$)} Given the type assignment $\gamma^{(s)}$ from the $s$-th iteration, construct indicator variables for each time $s$ and the assigned type $k$: $\mathbf{1}{\{t=s,k_i^{(s)}=k\}}$ for $s=-T_0 \cdots, -1$ and $k=1, \cdots, K$. By running OLS regression of $Y_{it} - Y_{it-1}$ on $X_{it}$ and the indicators, we update $\hat{\delta}_t^{(s)}(k)$ and $\hat{\theta}^{(s)}$.
    
    \item \textbf{(update $\gamma$)} Update $k_i^{(s)}$ for each $i$ by letting $k_i^{(s+1)}$ be the solution to the following minimization problem: for $i = 1, \cdots, N$, $$
    \min_{k\in \{1,\cdots,K\}} \sum_{t=-T_0}^{-1} \left( Y_{it} - Y_{it-1} - \hat{\delta}_t^{(s)} (k) - {X_{it}}^\intercal \hat{\theta}^{(s)}  \right)^2. 
    $$
     
    \item Repeat Step 1-2 until Step 2 does not update $\hat{\gamma}$, or some stopping criterion is met. For stopping criterion, one can set a maximum number of iteration or a minimum update in $\hat{\theta}^{(s)}$ and $\hat{\delta}^{(s)}$: set $S$ and $\varepsilon$ such that the iteration stops when $$
    s \geq S \hspace{3mm} \text{or} \hspace{3mm} \max \left\lbrace \left\| \hat{\theta}^{(s)} - \hat{\theta}^{(s-1)} \right\|_\infty, \left\| \hat{\delta}^{(s)} - \hat{\delta}^{(s-1)} \right\|_\infty \right\rbrace \leq \varepsilon.    
    $$
\end{enumerate}

In Appendix, we discuss Assumption \hyperlink{A7}{7} which extends Assumptions \hyperlink{separation}{4}-\hyperlink{A5}{5}. Under Assumption \hyperlink{A7}{7}, we have the following classification result. 

\begin{theorem}
\hypertarget{T2}{} Assumptions \hyperlink{finite}{3} and \hyperlink{A7}{7} hold. Then, up to some permutation on $\{1,\cdots,K\}$, $$
\Pr \left\lbrace \sup_i \mathbf{1}\{\hat{k}_i \neq k_i \} > 0 \right\rbrace = o\left( n {T_0}^{-\nu} \right) + o(1) \hspace{3mm} \forall \nu > 0
$$
as $n \to \infty$. 
\end{theorem}

\noindent \textit{Remark 3.} When $X_{it}$ is time-invariant, i.e. $X_{it} = X_i$, the linear model \eqref{eq:ParModel} and Assumption \hyperlink{A7}{7} can be understood as a parametric special case of the conditional parallel trend assumption: for $t < 0$, $$
\E \left[ Y_{it} (E_i) - Y_{it-1} (E_i) | k_i, X_i \right] = \delta_t(k_i) + {X_{i}}^\intercal \theta. 
$$

\noindent \textit{Remark 4.} Instead of assuming a linear structure on the first difference as in \eqref{eq:ParModel}, we can consider a linear model on the level of the outcome: \begin{align*}
Y_{it} (E_i)&= \alpha_i + \sum_{s=-T_0}^{t} \delta_s(k_i) + {X_{it}}^\intercal \theta + U_{it}, 
\intertext{and thus} 
Y_{it} (E_i)- Y_{it-1} (E_i)&= \delta_t(k_i) + \left( X_{it} - X_{it-1} \right)^\intercal \theta + U_{it} - U_{it-1}.
\end{align*}
The assumptions for the linear model in level is discussed in the Appendix along with Assumption \hyperlink{A7}{7}.

Theorem \hyperlink{T2}{2} finds the same rate on the misclassification probability as Theorem \hyperlink{T1}{1}. The key part of the proof utilizes the linear separability of $k_i$ and $X_{it}$. The proof firstly shows that $\theta$ is consistently estimated. Then, $Y_{it} - Y_{it-1} - {X_{it}}^\intercal \hat{\theta}$ is sufficiently close to $Y_{it} - Y_{it-1} - {X_{it}}^\intercal \theta$ so that the classification using $\hat{\theta}$ and the one using the true $\theta$ are the same. 

\subsection[Implementing treatment effect estimation with control covariates]{Implementing treatment effect estimation with $X_{it}$}

Theorem \hyperlink{T2}{2} implies that we can take the estimated types as given and apply the available treatment effect estimation methods when the rate given in Corollary \hyperlink{C1}{1} is satisfied. There are largely two ways to incorporate the control covariate $X_{it}$ in the treatment effect estimation. Firstly, we can follow an outcome model approach and impose a parametric model for the post-treatment outcome as we do for the pretreatment outcome in \eqref{eq:ParModel}. Given the parametric model, we plug in the estimated types as true types and estimate the model. A large variety of parametric models with a finite grouping structure can be used for the outcome model approach. For example, we can assume a linear model with type-specific coefficients for the treatment effect: for $t \geq 0$, \begin{align}
Y_{it} (E_i) = \alpha_i + \delta_t({k}_i) + \sum_{r \geq 0} \beta_r({k}_i) \mathbf{1}\{t = E_i + r\} + X_{it} ^\intercal \theta + U_{it}. \label{eq:ParLinear}
\end{align}
A more discussion on the outcome model approach is discussed in Section \hyperlink{Aom}{C.3} of the Appendix. The outcome model approach has the merit of developing  a parsimonious model for treatment effect; using the outcome model approach, we can impose some structure over how the latent type $k_i$ and the observable information $X_{it}$, $E_i$ interact in terms of the treatment effect heterogeneity.

Alternatively, we can use an assignment model approach. In the assignment model approach, instead of imposing a parametric model for the post-treatment outcomes, we impose a parametric model for the treatment timing. Suppose that we are given a time-invariant control covariate $X_i$ and that the conditional parallel trend assumption holds with $X_i$: for every $t, s \geq -1$, $$
\E \left[ Y_{it}(\infty) - Y_{is}(\infty) | k_i, X_i, E_i \right] = \E \left[ Y_{it}(\infty) - Y_{is} (\infty) | k_i, X_i \right].
$$ 
Then, we can apply the results of \citet{CS} by assuming an assignment model and estimating the propensity to be treated given the type $k_i$ and the control covariate $X_i$. For example, when $E_i \in \{0,\infty\}$, the logistic model can be used: $$
\Pr \left\lbrace E_i = 0 | k_i, X_i \right\rbrace = \frac{\exp \left( {X_i}^\intercal \theta + \delta(k_i) \right) }{1+\exp \left( {X_i}^\intercal \theta + \delta(k_i) \right)}.
$$
The benefit of the assignment model approach is that we allow for flexible interaction between the observable control covariates $X_i$ and the latent type $k_i$, in terms of the treatment effect heterogeneity. The assignment model approach is an extension of Corollary \hyperlink{C1}{1} since the type-specific diff-in-diff estimator defined in Section \hyperlink{Sec2}{2} is what we get when we assume the propensity score to be a trivial function of $X_i$: $\Pr \left\lbrace E_i = e | k_i, X_i \right\rbrace = \Pr \left\lbrace E_i = e | k_i \right\rbrace$. A more discussion on the assignment model approach is discussed in Section \hyperlink{Aam}{C.4}. 

\section{Simulation}

\hypertarget{Sec6}{}

In this section, we present simulation results to discuss the finite-sample performance of the type-specific diff-in-diff estimator, compared to some existing estimators in the literature. For that, we constructed a random sample using the following data generating process: for $t = -T_0-1, \cdots, 0$, \begin{align*}
Y_{it} &= \alpha_i + \delta(k) (t+1) + \beta(k_i) D_{i} \mathbf{1}\{t=0\} + U_{it}, \\
U_{it} &= \rho U_{it-1} + V_{it}.
\end{align*}
$D_i, \alpha_i, U_{i,-T_0-1}, \{V_{it}\}_{t \leq 0}$ are mutually independent given $k_i$. $D_i \big| k_i \sim \text{Bernoulli} \big(\pi(k_i) \big)$ and
\begin{align*}
\left( \alpha_i, U_{i,-T_0-1} \right) \big| k_i &\sim \mathcal{N} \left( \begin{pmatrix} \alpha(k_i) \\ 0 \end{pmatrix}, \begin{pmatrix} 17 & 0 \\ 0 & \sigma \end{pmatrix} \right), \\
V_{it} \big| k_i &\overset{iid}{\sim} \mathcal{N} \Big( 0,\sigma^2 ( 1-\rho^2 ) \Big).
\end{align*}
The values of the DGP parameters that pertain the classification step are taken from the empirical moments of the dataset used in the next section: $\sigma = 1.85$ and $\rho = 0.60$ for the error distribution and $\min_{k \neq k'} | \delta(k) - \delta(k') | = 1.32$ for the type separation.\footnotemark \footnotetext{The rest of the simulation parameters are as follows. For a two-types DGP where $K=2$, we set $\left( \pi(1), \pi(2) \right) = \left(1/3, 2/3 \right)$, $\left( \alpha(1), \alpha(2) \right) = \left(37,39\right)$, $\left( \delta(1), \delta(2) \right) = \left( 1.66, 0 \right)$ and $\left( \beta(1), \beta(2) \right) = \left( 4, 1 \right)$. $\Pr \left\lbrace k_i = 1 \right\rbrace = \Pr \left\lbrace k_i = 2 \right\rbrace=1/2$. For a three-types DGP where $K=3$, we set $\left( \pi(1), \pi(2), \pi(3) \right) = \left(1/3, 1/2, 1/2 \right)$, $\left( \alpha(1), \alpha(2), \alpha(3) \right) = \left(37,39,35 \right)$, $\left( \delta(1), \delta(2), \delta(3) \right) = \left( 2.74, 1.42, 0 \right)$ and $\left( \beta(1), \beta(2), \beta(3) \right) = \left( 5, 1, 0 \right)$. $\Pr \left\lbrace k_i = 1 \right\rbrace = \Pr \left\lbrace k_i = 2 \right\rbrace = 2/5$ and $\Pr \left\lbrace k_i = 3 \right\rbrace = 1/5$. The parameters concerning classification\textemdash $\beta, \sigma, \rho$\textemdash  are taken from the dataset used in Section \hyperlink{Sec7}{7}.} Note that a simple mean comparison of the treated units and untreated units is a biased estimator for the treatment effect when $\pi(k)$ is not constant in $k$. 

In the classification step, two different specifications for the type-specific time trend $\delta_t(k)$ were used. Firstly, we used the most flexible specification where $\delta_t(k)$ is allowed to vary across every $t$: $\{\delta_t(k)\}_{t\leq 0}$. Secondly, we assumed that the researcher has a priori knowledge on the DGP and imposes a constant slope restriction $\delta_t(k) = \delta_{t'}(k)$ for every $t,t'$, estimating only one parameter for each type: $\delta(k)$. Given the type classifications, we estimated the ATT using the type-specific diff-in-diff estimators; for comparison, we considered the unconditional diff-in-diff, the synthetic control and the synthetic diff-in-diff estimators.

Table \ref{tab:sim1} and Table \ref{tab:sim2} contain the simulated bias and the simulated MSE from 500 random samples. Also, they contain the type classification success rate: the first and the second columns of the bottom panel denote probability of a random sample having less than 5\% misclassification and the third and the fourth columns denote probability of perfect classification. For large $T_0=20,30$, both the type-specific diff-in-diff estimator and the synthetic diff-in-diff estimator perform well since there are many pretreatment outcomes to be used to control for the unit-level heterogeneity. However, in the case of the two-types DGP, the type-specific diff-in-diff estimator outperforms the other estimators for small $T_0=10$ since it best reflects the finite type structure in dataset. As for the classification, long pretreatments ($T_0=20, 30$) always achieves perfect classification, except when no smoothness restriction is imposed to the three-types DGP; even in that case, perfect classification probability is 0.80 for $T_0=20$ and 0.94 for $T_0=30$. 

In addition to the simulation specifications discussed above, we consider two additional DGPs to discuss misspecification in classification: firstly, we let $K=5$; secondly, we let the latent type be continuous. For both DGPs, we applied the two-types type-specific diff-in-diff estimator; the first DGP is a case of a misspecified number of types and the second DGP is a case of weakly separated types. The type-specific diff-in-diff estimator is more robust to the first type of misspecification, where the types are still strongly separated: for more discussion, see Section A.2.3. of the Supplementary Appendix.  

\section{Application}

\hypertarget{Sec7}{}

To show how the type-specific diff-in-diff estimator applies to a real dataset, we revisit \citet{L}. Since the Supreme Court ruling on Brown v. Board of Education of Topeka in 1954 that found state laws in the United States enabling racial segregation in public schools unconstitutional, various efforts have been made to desegregate public schools, including court-ordered desegregation plans. After several decades, another important Supreme Court case was made in 1991; the ruling on Board of Education of Oklahoma City v. Dowell in 1991 stated that school districts could terminate the court-ordered plans once it successfully removed the effects of the segregation. Since the second Supreme Court ruling, school districts started to file for dismissal of court-ordered desegregation plans, mostly in southern states.  

\citet{L} used the variation in timing of the district court rulings on the desegregation plan to estimate the effect on racial composition and education outcomes in public schools. The paper uses annual data on mid- and large-sized school districts from 1987 to 2006, obtained from the Common Core of Data (CCD), which contains data on school districts from 1987 to 2006, and the School District Databook (SDDB) of the US census, which contains data on school districts in 1990 and in 2000. To document if a school district was under a court-ordered desegregation plan at the time of the Supreme Court ruling in 1991 and when and if the school district got the desegregation plan dismissed at the district courts, \citet{L} collected data from various published and unpublished sources, including a survey by Rosell and Armor (1996) and the Harvard Civil Rights Project. 

Though \citet{L} looks at several outcome variable, we focus on one outcome variable, the dissimilarity index: the dissimilarity index for school district $i$ is  \begin{align*}
    Y_i &= \frac{1}{2} \sum_{j \in J_i} \left| \frac{b_{j}}{B_i} - \frac{w_{j}}{W_i}\right|  \times 100, \\
    b_{j}&: \text{ \# of black students in school } j , \hspace{5mm} w_{j} : \text{ \# of white students in school } j \\
    J_{i} &: \text{ the set of school in school district } i, \\
    B_i &= \sum_{j \in J_i} b_{j}, \hspace{5mm} W_i = \sum_{j \in J_i} w_{j}, \hspace{5mm} 
\end{align*}
The dissimilarity index ranges from 0 to 100, with 100 being perfectly segregated schools and 0 being perfectly representative schools.\footnotemark \footnotetext{In \citet{L}, the dissimilarity index ranges from zero to one but we rescaled the index for more visibility.} 

We followed the data cleaning process in the paper and chose the timespan of 1988-2007 to form a balanced panel of school districts that were under a court-ordered desegregation plan in 1988-1999, which gave us 50 school districts. We use the following linear model for the pretreatment outcomes: for $t=1989,\cdots,1999$,$$
Y_{it} (E_i) - Y_{it-1}(E_i) = \delta_t(k_i) + {X_{it}}^\intercal \theta + U_{it}. 
$$
The effective number of pretreatment outcomes is 11. The control covariates $X_{it}$ contain a central city indicator variable, percentage of students who are white, percentage of students who are hispanic, percentage of students with free/reduced-priced lunch and number of students. For the purpose of comparison, here we present the main empirical specification of \citet{L}:  \begin{align*}
    Y_{it} - Y_{it-1} = \delta_{jt} + \sum_{r=-4}^{10} \beta_r \left( \mathbf{1}\{t=E_i+r\} - \mathbf{1}\{t=E_i+r+1\} \right) + {X_i}^{\intercal} {\theta_t} + U_{it} 
\end{align*}
Though two specifications look alike, there are some differences. Firstly, though \citet{L} and we use the same control covariates, \citet{L} only used their values from the first year, with time-varying coefficient $\theta_t$: $X_i = X_{i,-T_0-1}$.\footnotemark \footnotetext{Also, \citet{L} used three additional variables: squared number of students, cubed number of students and squared percentage of students with free/reduced-price lunch.} On the other hand, we use time-varying control covariates $X_{it}$, with time-invariant coefficient $\theta$. Secondly, \citet{L} uses time fixed-effects $\delta_{jt}$ based on census region, which assigns every school district into one of the four regions. In the terminology of the model used in this paper, \citet{L} took the census region as the true type assignment whereas we used the data to estimate the type assignment. Lastly, the regression specification in \citet{L} has a dynamic treatment effect $\beta_r$ whereas we only impose linearity on pretreatment outcomes and therefore do not have any treatment effect term. Since $n=50$ is relatively small, we imposed an additional smoothness restriction on the type-specific time fixed-effects: $\delta_t(k) = \delta(k)$.\footnotemark \footnotetext{The constant slope restriction was chosen out of five specifications\textemdash constant slope, constant slope with a break, constant slope with two breaks, linear, linear with one break\textemdash, based on cross-validated mean-squared forecasting error. For more discussion, see the Supplementary Appendix.} Then, we applied the $K$-means clustering classifier with $K=2$.\footnotemark \footnotetext{As robustness check, we also considered $K=3$ and $K=4$. The Bayesian information criterion selected $K=3$ and the qualitative result remains the same for both $K=2$ and $K=3$. For more discussion, see Supplementary Appendix.} The first-step classification assigns 8 treated units and 14 never-treated units to Type 1 and 13 treated units and 15 never-treated units to Type 2; Type 2 school districts are slightly more likely to be treated. 

Given the first-step classification result, we conducted a within-type balancedness test; Table \ref{tab:withintypebal} contains within-type balancedenss test results using control covariates from $t=1988$. Within the two types, the control covariates are well-balanced across treatment status: treated v. never-treated. Thus, we did not use $X_{it}$ in the estimation step and use the unconditional type-specific diff-in-diff estimator as defined in \eqref{eq:dCATTest}; the target parameter captures the type-specific dynamic ATT that assigns equal weights to each treated unit within the same type. 

\begin{figure}[p]
\centering
\textsc{\caption{Type-specific ATT} \label{fig:K2}} 
\vspace{3mm} 
\begin{tikzpicture}
{
\draw[->, line width=1] (-4.5*1.2,0) -- (4.5*1.2,0) node[anchor=west] {$r$};
\draw[->, line width=1] (-4.5*1.2,-3) -- (-4.5*1.2,5.5) node[anchor=south] {$\beta_r(k)$}; 

\draw[dotted, gray] (-4*1.2,-3) -- (-4*1.2,5.5);
\draw[dotted, gray] (-3*1.2,-3) -- (-3*1.2,5.5);
\draw[dotted, gray] (-2*1.2,-3) -- (-2*1.2,5.5);
\draw[dotted, gray] (-1*1.2,-3) -- (-1*1.2,5.5);
\draw[dotted, gray] (0*1.2,-3) -- (0*1.2,5.5);
\draw[dotted, gray] (1*1.2,-3) -- (1*1.2,5.5);
\draw[dotted, gray] (2*1.2,-3) -- (2*1.2,5.5);
\draw[dotted, gray] (3*1.2,-3) -- (3*1.2,5.5);
\draw[dotted, gray] (4*1.2,-3) -- (4*1.2,5.5);

\node[anchor=north] at (-4*1.2,-0.4) {-4};
\node[anchor=north] at (-3*1.2,-0.4) {-3};
\node[anchor=north] at (-2*1.2,-0.4) {-2};
\node[anchor=north] at (-1*1.2,-0.4) {-1};
\node[anchor=north] at (0*1.2,-0.4) {0};
\node[anchor=north] at (1*1.2,-0.4) {1};
\node[anchor=north] at (2*1.2,-0.4) {2};
\node[anchor=north] at (3*1.2,-0.4) {3};
\node[anchor=north] at (4*1.2,-0.4) {4};

\node[anchor=east] at (-4.5*1.2,-2.5) {-5};
\node[anchor=east] at (-4.5*1.2,0) {0};
\node[anchor=east] at (-4.5*1.2,2.5) {5};
\node[anchor=east] at (-4.5*1.2,5) {10};
}

{    
\draw[dashed, line width=1, red] (-4*1.2,-2.8730723*0.5) -- (-3*1.2,-2.1279914*0.5) -- (-2*1.2,-1.1975863*0.5) -- (-1*1.2,0) -- (0*1.2,-0.5244448*0.5) -- (1*1.2,-0.6530800*0.5) -- (2*1.2,0.7120334*0.5) -- (3*1.2,1.6235811*0.5) -- (4*1.2,3.2324551*0.5); 
             
\draw[dashed, line width=1, red] (-4*1.2,0.77334657*0.5) -- (-3*1.2,0.46655613*0.5) -- (-2*1.2,0.08292767*0.5) -- (-1*1.2,0) -- (0*1.2,3.33358039*0.5) -- (1*1.2,6.42063475*0.5) -- (2*1.2,8.34846424*0.5) -- (3*1.2,10.00665698*0.5) -- (4*1.2,11.31837804*0.5); 

\draw[line width=2, red] (-4*1.2,-1.04986284*0.5) -- (-3*1.2,-0.83071762*0.5) -- (-2*1.2,-0.55732934*0.5) -- (-1*1.2,0) -- (0*1.2,1.40456778*0.5) -- (1*1.2,2.88377739*0.5) -- (2*1.2,4.53024881*0.5) -- (3*1.2,5.81511902*0.5) -- (4*1.2,7.27541656*0.5) node[anchor=west] {type 1}; 
}

{
\draw[dashed, line width=1, blue] (-4*1.2,-1.4305809*0.5) -- (-3*1.2,-1.5170373*0.5) -- (-2*1.2,-0.7211843*0.5) -- (-1*1.2,0) -- (0*1.2,-1.2006542*0.5) -- (1*1.2,-0.7837984*0.5) -- (2*1.2,-1.4606200*0.5) -- (3*1.2,-2.4710323*0.5) -- (4*1.2,-2.3986732*0.5); 

\draw[dashed, line width=1, blue] (-4*1.2,1.97878593*0.5) -- (-3*1.2,1.02261507*0.5) -- (-2*1.2,0.53894866*0.5) -- (-1*1.2,0) -- (0*1.2,0.53715646*0.5) -- (1*1.2,3.97343603*0.5) -- (2*1.2,5.39230923*0.5) -- (3*1.2,6.96411762*0.5) -- (4*1.2,7.55825494*0.5) ; 
       
\draw[line width=2, blue] (-4*1.2,0.27410250*0.5) -- (-3*1.2,-0.24721113*0.5) -- (-2*1.2,-0.09111783*0.5) -- (-1*1.2,0) -- (0*1.2,-0.33174886*0.5) -- (1*1.2,1.59481883*0.5) -- (2*1.2,1.96584461*0.5) -- (3*1.2,2.24654268*0.5) -- (4*1.2,2.57979085*0.5) node[anchor=west] {type 2}; 
}

{
\draw[dashed, line width=1, Green] (-4*1.2,-0.6251*0.5) -- (-3*1.2,-1.1220*0.5) -- (-2*1.2,-0.5295*0.5) -- (-1*1.2,-0.5548*0.5) -- (0*1.2,-0.9898*0.5) -- (1*1.2,-0.4420*0.5) -- (2*1.2,-0.2939*0.5) -- (3*1.2,-0.4986*0.5) -- (4*1.2,-0.1881*0.5); 

\draw[dashed, line width=1, Green] (-4*1.2,0.9210*0.5) -- (-3*1.2,0.6863*0.5) -- (-2*1.2,1.5395*0.5) -- (-1*1.2,1.3804*0.5) -- (0*1.2,1.6585*0.5) -- (1*1.2,4.7930*0.5) -- (2*1.2,6.7187*0.5) -- (3*1.2,8.3733*0.5) -- (4*1.2,9.8802*0.5) ; 
       
\draw[line width=2, Green] (-4*1.2,0.1480*0.5) -- (-3*1.2,-0.2178*0.5) -- (-2*1.2,0.5050*0.5) -- (-1*1.2,0.4128*0.5) -- (0*1.2,0.3343*0.5) -- (1*1.2,2.1755*0.5) -- (2*1.2,3.2124*0.5) -- (3*1.2,3.9374*0.5) -- (4*1.2,4.8461*0.5) node[anchor=west] {CSA}; 
}

\end{tikzpicture}
\vspace{2mm} \\
    \begin{minipage}{0.7\textwidth}
    \footnotesize
    The graph reports estimates for the dynamic effect of dismissing court-mandated desegregation plan on the dissimilarity index of a school district. The dissimilarity index ranges from 0 to 100. In 1988, the average dissimilarity index was 34 and the standard deviation was 16. 
    
    \vspace{2mm} 
    
    The red line and the blue line are type-specific diff-in-diff estimates for the type-specific dynamic ATT. Type 1 is the type where the dissimilarity index was rising faster and Type 2 is the type where the dissimilarity index was rising slower. The green line is the estimates from \citet{CS}; for \citet{CS}, only the central city indicator and the percentage of students who are white in 1988 are used, to avoid multicollinearity.
    
    \vspace{2mm} 
    
    The dashed lines denote the confidence intervals are at 0.05 significance level and are computed with asymptotic standard errors. 
    \end{minipage}
\end{figure}
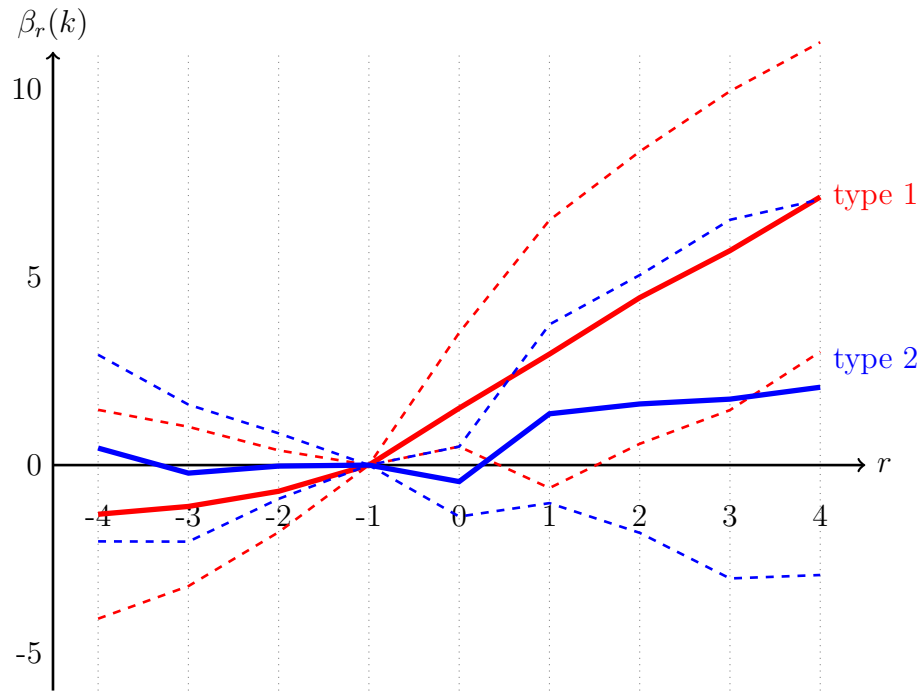

Figure \ref{fig:K2} contains the type-specific diff-in-diff estimates for Type 1 and Type 2 school districts. From Figure \ref{fig:K2}, we see that the treatment effect is bigger for Type 1 and smaller for Type 2; the termination of court-ordered desegregation plans exacerbated racial segregation more severely for Type 1. The type-specific estimation provide a sensible aggregate estimate; averaging the type-specific diff-in-diff estimates across types gives us $(8 \hat{\beta}_r(1) + 13 \hat{\beta}_r(2))/21  = 4.37$ at $r=4$ whereas the pooled regression estimate from \citet{L} is around 4-5 at $r=4$, depending on specifications, and the conditional diff-in-diff estimate from \citet{CS} is 4.85 at $r=4$. For reference on the magnitude, the mean of the dissimilarity index was 34 and its standard deviation was around 16 in 1988. Also, Figure \ref{fig:K2} contains estimates for $\beta_r(k)$ such that $r < 0$; none of the pretreatment trend was found to be away from zero at 0.05 significance level. 

So, estimates on treatment effect suggest that Type 1 and Type 2 are different; the Type 1 school districts are more responsive to the treatment. How are these types different in other regards? Firstly, Table \ref{tab:bal} shows us some descriptive statistics on the outcome variable and other control covariates for each type, using year 1988 data. The null hypothesis that the entire vector of mean differences between Type 1 and Type 2 is zero is rejected with a $t$-test at size 0.05; the Type 1 school districts are different from the Type 2 school districts in terms of their observable characteristics. For instance, Type 1 school districts have higher proportion of white students and lower proportion of hispanic students. Secondly, in terms of the unobserved heterogeneity captured by the latent type variable, Type 1 has seen a steeper increase in the dissimilarity index while the slope was smaller for Type 2: $\Big( \hat{\delta}(1), \hat{\delta}(2) \Big) = (3.59, 1.93)$. This implies that the dismissal of desegregation plans had a bigger impact on Type 1, where the dissimilarity index was already rising faster. This observation presents future research questions: for example, why do the school districts that were getting more segregated also get affected more from the dismissal of the desegregation plan? 

\section{Conclusion}

In this paper, we introduce a type-specific parallel trend assumption in a panel data model with a latent type. By assuming the latent type variable has a finite support and is well-separated in the long pretreatment time series, the $K$-means classifier estimates the true types consistently. Also, based on the estimated types, we estimate the type-specific treatment effect. The type-specific diff-in-diff estimator is useful when we suspect heterogeneity in time trends across units and want to explore the associated treatment effect heterogeneity. By applying the estimation method to an empirical application, we find some interesting empirical results where the estimates on the type-specific treatment effects and those on the type-specific time trend tell a story: the effect of terminating court-mandated desegregation plans were bigger for school districts where the dissimilarity index was growing. 

\bibliographystyle{aer}
\bibliography{mylit}

\pagebreak

\appendix

\begin{center}
    {\Large APPENDIX}
\end{center}

\section{Parallel trend v. design-based approach}

\hypertarget{Asec1}{}

The type-specific parallel trend assumption used in this paper do not impose restrictions on the assignment process for the treatment timing and rather directly impose restrictions on the outcome model. Though the parallel trend type assumptions have their own advantages of being concise and straightforward, the parallel trend assumption hinges on an arbitrary choice of what to compare: the temporal differences in level. For example, when a researcher is interested in estimating the treatment effect as a percentage change of the outcome variable, they may be motivated use a parallel trend assumption with logged outcome variables: $$
\E \left[ \log Y_{it}(\infty) - \log Y_{is}(\infty) | k_i, E_i \right] = \E \left[ \log Y_{it}(\infty) - \log Y_{is}(\infty) | k_i \right].
$$
\noindent On the other hand, a design-based approach such as a unconfoundedness assumption would be free of this commitment to a functional form. When \begin{align}
\left\lbrace Y_{it}(e) \right\rbrace_{t,e} \independent E_i | k_i, \label{eq:uncon}
\end{align}
a parallel trend assumption with any functional form would hold.\footnotemark \footnotetext{This statement is only true for the unconfoundedness assumption as given in \eqref{eq:uncon}. \citet{DL} discuss a simple two period case ($t=1,2$) where no one is treated at $t=1$ and show that the parallel trend assumption and the unconfoundedness assumption do not nest each other when the unconfoundedness assumption is applied sequentially: $Y_{i1}(\infty) = Y_{i1}(2)$ and $Y_{i2}(\infty) \independent D_{i2} | \big( k_i, Y_{i1} \big)$} This comes at a cost of assuming distributional independence, which is stronger than the mean independence used in the parallel trend type assumption. 

There are some benefits to the design-based approach in addition to being robust to the choice of functional form. Under the parallel trend type assumption, there was no clear choice in which temporal differences to use. However, when we assume random treatment timing or the unconfoundedness assumption, we have some theoretical guidance on this choice. \citet{RSjpem} assume random treatment timing and find an efficient estimator among diff-in-diff type estimators that uses different weights across different temporal differences. Given the same classification result from Theorem \hyperlink{T1}{1}, the unconfoundedness assumption \eqref{eq:uncon} can be used to find an efficient type-specific diff-in-diff estimator following the procedure of \citet{RSjpem}, using the same argument from the proof for Corollary \hyperlink{C2}{2}: the classification error is faster than $1/\sqrt{n}$.

When a researcher does choose to follow a design-based approach, the question of great interest is how much more restrictions are imposed when assuming the unconfoundedness, compared to the conditional parallel trend. \citet{RSecma,GSW} provide insights to this question. \citet{RSecma} show that an equivalent condition for the parallel trend assumption to hold for any monotone transformation of the outcome variable is that the population is divided into two subgroups where the treatment is random for the first subgroup and the untreated potential outcome has time-invariant distribution for the second subgroup. In this sense, the unconfoundedness assumption \eqref{eq:uncon} is indeed strictly stronger than the type-specific parallel trend assumption holding for every monotone transformation of the outcome. \citet{GSW} provide insight in understanding the cost of assuming an additional parallel trend type assumption incrementally. Given a functional form, \citet{GSW} provide  necessary conditions and sufficient conditions for that specific parallel trend assumption in terms of restrictions on the assignment model.

\section{Weighted outcomes as counterfactual}

In this section, we discuss how the type-specific diff-in-diff estimator compares to other treatment effect estimators that assign weights over control units to construct a counterfactual outcome. In specific, we consider the conventional diff-in-diff estimator and the synthetic control estimator. 

Find that the classification result from \eqref{eq:estimator} satisfy that $$
\hat{\delta}_t(k) = \frac{\sum_{i=1}^n \left( Y_{it} - Y_{it-1} \right) \mathbf{1}\{\hat{k}_i=k\}}{\sum_{i=1}^n \mathbf{1}\{\hat{k}_i=k\}}.
$$
$\hat{\delta}_t(k)$ puts equal weights over $Y_{it} - Y_{it-1}$ for units with the same estimated type $k$. Consider a simple case where there is only one post-treatment period and only one treated unit: $T_1=1$, $E_i = \infty$ for every $i \leq N_0=n-1$ and $E_n = 0$. Consider a treatment effect estimator $\hat{\beta}$ which can be written as a weighted sum of $Y_{it}$: $\hat{\beta} = \sum_{i,t} w_{it} Y_{it}$. In a simple diff-in-diff estimator using $t \in \{-1,0\}$, the weight is \begin{align*}
w_{it}^{did} &= \mathbf{1}\{i=n,t=0\} - \mathbf{1}\{i=n,t=-1\} - \frac{\mathbf{1}\{i \leq N_0,t=0\}}{N_0} + \frac{\mathbf{1}\{i \leq N_0,t=-1\}}{N_0}.
\end{align*}
In the case of the synthetic control (see \citet{ADH10,ADH15}, $$
w_{it}^{sc} = \mathbf{1}\{i=n,i=0\} - \sum_{j=1}^{N_0} w_j^* \mathbf{1}\{i=j,t=0\}
$$
where $\{w_j^*\}_{j \leq N_0}$ are solution to the following minimization: $$
\min_{w} \sum_{t=-T_0-1}^{-1} \left( Y_{nt} - \sum_{i=1}^{N_0} w_{i} Y_{it} \right)^2.
$$
subject to $\sum_{i=1}^{N_0} w_{i} = 1$ and $w_{i} \geq 0$. In the case of the type-specific diff-in-diff, $$
w_{it}^{tdid} =  \mathbf{1}\{i=n,t=0\} - \mathbf{1}\{i=n,t=-1\} - \sum_{j=1}^{N_0} w_j^{**} \Big( \mathbf{1}\{i=j,t=0\} - \mathbf{1}\{i=j,t=-1\} \Big)
$$
where $\{w_j^{**}\}_{j \leq N_0}$ are (a function of) the solution to the following minimization: $$
\min_w \sum_{i=1}^{n} \sum_{t=-T_0}^{-1} \left( \left( Y_{it} - Y_{it-1} \right) - \sum_{j} w_{ij} \left( Y_{jt} - Y_{jt-1} \right) \right)^2 
$$
subject to $w_{ij} = \mathbf{1}\{k_i=k_j\} \big/ \sum_{l} \mathbf{1}\{k_l=k_i\}$ for some $\{k_i\}_{i=1}^n \in \{1,\cdots,K\}^n$. Based on the optimized $w_{ij}$, we get $w_{j}^{**} = w_{nj} \big/ \sum_{j' \neq n} w_{nj'}$. 

Compared to the diff-in-diff estimator, the type-specific diff-in-diff estimator admits more flexible cross-sectional weights by possibly using only a subset of the never-treated units. Compared to the synthetic control estimator, the type-specific diff-in-diff estimator is less flexible in terms of the cross-sectional weights since it is dichotomous cross-sectionally; a never-treated unit gets a uniform weight if and only if it shares the same type with the treated unit and gets zero weight otherwise. However, the synthetic control estimator assigns nonzero weights only to contemporaneous outcomes while the type-specific diff-in-diff estimator takes temporal difference. Lastly, though the weights are not as easily formulated as with other methods discussed here, the synthetic diff-in-diff estimator from \citet{Aetal} also uses a weighted sum type estimator and assigns flexible weights both cross-sectionally and intertemporally, therefore nesting all of the methods discussed above.

\section{Additional assumptions and asymptotic results}

\subsection{Overlap condition}

\hypertarget{Aol}{}

\begin{assumption}
\hypertarget{A6}{} Fix some $t \in \{0, \cdots, T_1-1\}$, $k \in \{1,\cdots, K\}$, $e \in \{0,\cdots, T_1-1\}$ and $r \in \{0,\cdots, T_1-1\}$. \begin{enumerate}
    \item[\textbf{a.}] $\Pr \left\lbrace E_i=e | k_i=k \right\rbrace \cdot \Pr \left\lbrace E_i > t | k_i =k \right\rbrace >0$.
    \item[\textbf{b.}] $\Pr \left\lbrace E_i=e' | k_i=k \right\rbrace \cdot \Pr \left\lbrace E_i > e'+r | k_i =k \right\rbrace >0$ for some $e' \leq T_1 - 1 - r$.
\end{enumerate}
\end{assumption}

Assumption \hyperlink{A6}{6-a} assumes nonzero treatment probability at time $e$ and at some time later than $t$, for a unit with type $k$. In Lemma \hyperlink{L1}{1} and Corollary \hyperlink{C1}{1}, Assumption \hyperlink{A6}{6-a} is the overlap condition for $ATT_t(k,e)$. 

Assumption \hyperlink{A6}{6-b} assumes there exists two treatment timings $e' \in \{0,\cdots, T_1-1-r\}$ and $s \in \{e'+r+1,\cdots, T_1-1, \infty \}$ with nonzero treatment probability for a unit with type $k$; i.e., the treatment timings of the two treatment cohorts $\{i:E_i=e'\}$ and $\{i:E_i=s\}$ differ by at least $r$ and both of the treatment cohorts have nonzero measure. For Assumption \hyperlink{A6}{6-b}, it helps to think of some $\bar{r}_k \geq 0$ such that $$
\bar{r}_k = \min \left\lbrace T_1, \max \left\lbrace e: \Pr \left\lbrace E_i=e|k_i=k \right\rbrace >0 \right\rbrace \right\rbrace - \min \left\lbrace e: \Pr \left\lbrace E_i=e | k_i=k \right\rbrace > 0 \right\rbrace - 1.
$$
$\bar{r}_k$ is the upper bound on how far the dynamic treatment effect can be identified; for any $r \leq \bar{r}_k$, the dynamic ATT $\beta_r(k)$ is identified. In Lemma \hyperlink{L1}{1} and Corollary \hyperlink{C1}{1}, Assumption \hyperlink{A6}{6-b} is the overlap condition for $\beta_r(k)$. 

\subsection[Type classification given control covariates]{Type classification given control covariates $X_{it}$}

\hypertarget{At}{}

To have the classification result under the linear outcome model with control covariates \eqref{eq:ParModel}, we assume the following assumption. 
\begin{assumption}
\hypertarget{A7}{} 
With some $M, \tilde{M} >0$,
\begin{itemize}
    \item[\textbf{a.}] (iid across units) $\left( \left\lbrace X_{it}, U_{it} \right\rbrace_{t < 0}, E_i, k_i \right) \overset{\text{iid}}{\sim} F$. 
    
    \item[\textbf{b.}] (compact parameter space) For every $t$ and $k$, $| \delta_t(k)| \leq M$. $\| \theta \|_2 \leq M$. 

    \item[\textbf{c.}] (well-separated types) Whenever $k \neq k'$, $$
    \frac{1}{T_0}\sum_{t=-T_0}^{-1} \left( \delta_t(k) - \delta_t(k') \right)^2 \to c(k,k') > 0
    $$
    as $n \to \infty$. 
    
    \item[\textbf{d.}] (strict exogeneity and finite moments) \\
    For every $t <0 $, $\E \big[ U_{it} | k_i, \left\lbrace X_{is} \right\rbrace_{s=-T_0-1}^{-1} \big]= 0$ and $\E \big[ {U_{it}}^4 | k_i, \left\lbrace X_{is} \right\rbrace_{s=-T_0-1}^{-1} \big] \leq M$. \\
    For every $t,s <0$, $\E \big[ \big| {X_{it}}^\intercal X_{is}  \big| \big] \leq M$. For any $\nu > 0$, $$
    \Pr \left\lbrace \frac{1}{T_0} \sum_{t=-T_0}^{-1} \| X_{it} \|_2 \geq \tilde{M} \right\rbrace = o \left( {T_0}^{-\nu} \right)
    $$
    as $n \to \infty$. 
    
    \item[\textbf{e.}] (long pretreatment) $T_0 \to \infty$ as $n \to \infty$. 

    \item[\textbf{f.}] (no measure zero types) For all $k \in \{1, \cdots, K\}$, $\Pr \left\lbrace k_i = k \right\rbrace > 0$
    
    \item[\textbf{g.}] (weakly dependent, thin-tailed errors) With some positive constant $d_1$ and $a$, $\left\lbrace U_{it} \right\rbrace_{t=-T_0}^{-1}$ is strongly mixing with mixing coefficient $\alpha[t]$ such that $\alpha[t] \leq \exp ( - a t^{d_1} )$. Also, with some positive constant $d_2$ and $b$, $U_{it}$ satisfies the following tail probability: for any $u > 0$,
    $$
    \Pr \left\lbrace \left| U_{it} \right| \geq u \right\rbrace \leq \exp \left( 1 - \left( u/b \right)^{d_2} \right)
    $$
    uniformly over $t < 0$.

    \item[\textbf{h.}] (no multicollinearity) Given an arbitrary type assignment $\tilde{\gamma} = \left( \tilde{k}_1, \cdots, \tilde{k}_n \right)$, let $\bar{{X}}_{k \wedge \tilde{k},t}$ denote the mean of $X_{it}$ among units such that $k_i=k$ and $\tilde{k}_i=\tilde{k}$. Let $\rho_{n}(\tilde{\gamma})$ denote the minimum eigenvalue of the following matrix: $$
    \frac{1}{n T_0} \sum_{i=1}^n \sum_{t=-T_0}^{-1} \left( X_{it} -  \bar{{X}}_{k_i \wedge \tilde{k}_i,t} \right) \left( X_{it} - \bar{{X}}_{k_i \wedge \tilde{k}_i,t} \right)^\intercal.
    $$
    Then, $\min_{\tilde{\gamma} \in \Gamma} \rho_n(\tilde{\gamma}) \xrightarrow{p} \rho$ as $n \to \infty$. 
\end{itemize}
\end{assumption}
Assumption \hyperlink{A7}{7-c} replaces Assumption \hyperlink{separation}{4} in the context of \eqref{eq:ParModel}. Assumption \hyperlink{A7}{7-c} assumes that the residual unobserved heterogeneity across units after regressing out $X_{it}$ has finite types and is well-separated in the $l_2$ norm. Assumption \hyperlink{A7}{7-d} replaces Assumption \hyperlink{A5}{5-b} and additionally assumes that for large enough $\tilde{M}$,  the probability of $\frac{1}{T_0} \sum_{t=-T_0}^{-1} \| X_{it} \|_2$ being larger than $\tilde{M}$ goes to zero exponentially. Moreover, Assumption \hyperlink{A7}{7-d} combined with \eqref{eq:ParModel} replaces the parallel trend assumption given in Assumptions \hyperlink{A1}{1}-\hyperlink{A2}{2}, by imposing $$
\E \left[ Y_{it} - Y_{it-1} - {X_{it}}^\intercal \theta | k_i, \{X_{is}\}_{s=-T_0-1}^{-1} \right] = \delta_t(k_i).
$$
Assumption \hyperlink{A7}{7-g} replaces Assumption \hyperlink{A5}{5-e}. Assumption \hyperlink{A7}{7-h} assumes that there is sufficient variation in $X_{it}$ within each type. When an outcome model is assumed for the pretreatment outcome in level as in Remark 4, the same conditions from Assumption \hyperlink{A7}{7-d,g} and an adjusted version of Assumption \hyperlink{A7}{7-h} by replacing $X_{it}$ with $X_{it} - X_{it-1}$ give us the same classification result as in Theorem \hyperlink{T2}{2}. 

\subsection[Outcome model approach given control covariates]{Outcome model approach given control covariates $X_{it}$}
\hypertarget{Aom}{} 

Once $\{k_i \}_{i=1}^n$ is estimated, we may use the estimated types to estimate various models on post-treatment periods with type-specific parameters. Directly modelling the outcome model with the observable information $X_{it}$ as in \eqref{eq:ParLinear} can be helpful when we are interested in treatment effect heterogeneity and we would like to impose some restrictions on the heterogeneity due to the structure of $X_{it}$. For example, when $X_{it}$ is continuous and multidimensional, a linearity assumption on the treatment effect $\beta_r(X_{it},k_i) = \beta_r(k_i)^\intercal X_{it}$ can be helpful in summarizing how $X_{it}$ interacts with the type $k_i$ in terms of the treatment effect. 

Consider a generalized outcome model for post-treatment outcomes: for $t \geq 0$, \begin{align}
    Y_{it}(E_i) - Y_{it-1}(E_i) &= m ( X_{it}, E_i, k_i; \xi) + U_{it}.\footnotemark  \notag
\end{align}
\noindent An important restriction we need to assume for the outcome model $m$ is that $m$ treats the latent type $k_i$ as a categorical variable; our type classifier only estimates a permutation on $\{1,\cdots,K\}$. In the example \eqref{eq:ParLinear}, $\xi = \Big( \{\delta_t(k)\}_{t \geq 0, k}, \{\beta_{r} (k) \}_{r\geq 0,k}, \theta \Big)$. Note that the dimension of $\xi$ is fixed; the dimension is $2T_1 K + p$ and $T_1$ and $K$ are fixed. Alternatively, we can consider a richer control specification: for example, $$
Y_{it} (E_i) = \alpha_i + \delta_t({k}_i) + \sum_{r \geq 0} \beta_r({k}_i)^\intercal X_{it} \mathbf{1}\{t = E_i + r\} + {X_{it}}^\intercal \theta(E_i) + U_{it}. 
$$
The treatment effect $\beta_r(k_i)^\intercal X_{it}$ depends on the type and the control covariate $X_{it}$ and the baseline trend $\delta_t(k) + {X_{it}}^\intercal \theta(E_i)$ depends on the type, the treatment cohort and the control covariates $X_{it}$. 

Let $\tilde{\xi}$ be the infeasible least-square estimator for $\xi$ and $\hat{\xi}$ be the plug-in least-square estimator for $\xi$: \footnotetext{Though the first-differenced outcome variables are used in the post-treatment outcome model for internal consistency with \eqref{eq:ParModel}, we can also consider models with outcome variable in level. In that case, one could use unit fixed-effects or type-by-treatment-cohort fixed-effects to address unit-level heterogeneity in level.} \begin{align*}
\tilde{\xi} &= \arg \min_{\xi \in \Xi} \frac{1}{n T_1} \sum_{i=1}^n \sum_{t=0}^{T_1-1} \Big( Y_{it} - Y_{it-1} - m(X_{it}, E_i, k_i; \xi ) \Big)^2, \\
\hat{\xi} &= \arg \min_{\xi \in \Xi} \frac{1}{n T_1} \sum_{i=1}^n \sum_{t=0}^{T_1-1} \left( Y_{it} - Y_{it-1} - m(X_{it}, E_i, \hat{k}_i; \xi ) \right)^2.
\end{align*}

\begin{assumption}
\hypertarget{A8}{} The latent type $k_i$ enters the outcome model $m$ as a categorical variable. $\Xi$, the parameter space for $\xi$, is bounded: with some $M> 0$, $$
\sup_{\xi \in \Xi} \| \xi \|_2 \leq M.
$$
The true value $\xi$ lies in the interior of $\Xi$. Also, the infeasible estimator $\tilde{\xi}$ satisfies that $$
\sqrt{n} \left( \tilde{\xi} - \xi \right) \xrightarrow{d} \mathcal{N} \left( \mathbf{0}, \Sigma \right)
$$
with some $\Sigma > 0$ as $n \to \infty$.
\end{assumption}

\begin{corollary}
\hypertarget{C2}{} Let Assumptions \hyperlink{finite}{3} and \hyperlink{A7}{7}-\hyperlink{A8}{8} hold. There exists some $\nu^*>0$ such that $n / {T_0}^{\nu^*} \to 0$ as $n \to \infty$. Then, up to some permutation on $\{1,\cdots,K\}$, $$
\sqrt{n} \left( \hat{\xi} - \xi \right) \xrightarrow{d} \mathcal{N} \left( \mathbf{0}, \Sigma \right)
$$
with $\Sigma > 0$ from Assumption \hyperlink{A8}{8} as $n \to \infty$.
\end{corollary}

\begin{proof}
The result holds directly from Theorem \hyperlink{T2}{2}. Find that \begin{align*}
    \sqrt{n} \left\| \tilde{\xi} - \hat{\xi} \right\|_2 &\leq 2\sqrt{n} M \mathbf{1} \left\lbrace \sup_i \mathbf{1}\{\hat{k}_i \neq k_i \} > 0 \right\rbrace = o_p(1)
\end{align*}
since for any $\varepsilon > 0$, \begin{align*}
\Pr \left\lbrace 2\sqrt{n} M \mathbf{1} \left\lbrace \sup_i \mathbf{1}\{\hat{k}_i \neq k_i \} > 0 \right\rbrace > \varepsilon \right\rbrace &\leq \Pr \left\lbrace \sup_i \mathbf{1}\{\hat{k}_i \neq k_i\} > 0\right\rbrace = o(1)
\end{align*}
from $n/{T_0}^{\nu^*} \to 0$ as $n \to \infty$.
\end{proof}

Note that Assumption \hyperlink{A8}{8} does not discuss whether the true parameter $\xi$ has sensible causal interpretation as we did for $ATT_t(k,e)$ in Section \hyperlink{Sec3}{3}. In the example of \eqref{eq:ParLinear}, it is well known that the linear coefficients $\beta_{r}(k)$ may suffer from the bias that comes from the dependence structure in $\mathbf{1}\{t=E_i+r\}$, given treatment effect heterogeneity.\footnotemark \footnotetext{It has been discussed that treatment effect estimators from TWFE specification are biased under the parallel trend type assumption (see \citet{CD,G,BJS,SA} among others) and potentially distort hypothesis testing (see \citet{BLW}). Also, \citet{GHK} show that even when the treatment timing is random, treatment effect estimators still suffer from contamination bias when dynamic treatment effect specification is used. } Thus, we consider an alternative approach in the next subsection. 

\subsection[Assignment model approach given control covariates]{Assignment model approach given control covariates $X_{it}$}
\hypertarget{Aam}{}

Directly modelling the outcome model may be too restrictive in some empirical contexts where the treatment effect depends on the control covariates $X_{it}$ and the type $k_i$ in a more flexible way. A similar concern is addressed in \citet{CS} where the authors consider a conditional parallel trend assumption where the conditioning set is the control covariates $X_{i}$. In \citet{CS}, authors impose restriction on the assignment model while not imposing any restriction on the treatment effect heterogeneity in terms of the control covariate $X_i$. 

With some finite-dimensional parameter $\xi$, we use a parametric function $\pi_e$ to model the conditional distribution of the treatment timing $E_i$ given the control covariate $X_i$ and the latent type $k_i$:\footnotemark \footnotetext{The conditional distribution of $E_i$ given $\left( k_i, X_i \right)$ captures how treatment timing depends on the type and the control covariate. However, it does not contain any information on the dependence between $k_i$ and $X_i$. For that end, we could consider the conditional distribution of $k_i$ given $X_i$. Given a new draw of $X_i$, we cannot know $k_i$; however, we can look at the (estimated) distribution of $k_i | X_i$. Moreover, based on the distribution, a prediction on treatment effect can also be made. A close parallel with the IV literature exists here; we cannot know if a newly drawn unit with covariate $X_i$ is a complier or not, but we can identify the conditional probability of them being a complier given $X_i$.} \begin{align*}
    \Pr \left\lbrace E_i = e| k_i, X_i \right\rbrace = \pi_e (X_i, k_i,\xi).
\end{align*}
Again, the type variable $k_i$ is treated as a categorical variable in the assignment model $\pi_e$. An example of such a model is an ordered logit model with type-specific coefficients: \begin{align*}
    \Pr \left\lbrace E_i \leq e | k_i=k, X_i=x \right\rbrace &= \frac{\sum_{e'=0}^{e} \exp (x^\intercal\theta(k) + \delta_{e'}(k))}{\sum_{e'=0}^{T_1 -1} \exp (x^\intercal\theta(k) + \delta_{e'}(k)) + \exp (x^\intercal\theta + \delta_\infty(k)) }.
\end{align*}
In this example, $\xi=\left( \theta, \delta_e(k) \right)_{k,e}$ and its dimension is fixed: $T_1 K + p$. 

Let $\tilde{\xi}$ be the infeasible maximum likelihood estimator for $\xi$ and $\hat{\xi}$ be the plug-in estimator for $\xi$: \begin{align*}
    \tilde{\xi} &= \arg \min_{\xi \in \Xi} \frac{1}{n} \sum_{i=1}^n \left( \sum_{e=0}^{T_1-1} \mathbf{1}\{E_i=e\} \log \pi_e (X_i,k_i,\xi) + \mathbf{1}\{E_i=\infty\} \log \pi_\infty (X_i,k_i,\xi) \right) , \\
    \hat{\xi} &= \arg \min_{\xi \in \Xi} \frac{1}{n} \sum_{i=1}^n \left( \sum_{e=0}^{T_1-1} \mathbf{1}\{E_i=e\} \log \pi_e (X_i,\hat{k}_i,\xi) + \mathbf{1}\{E_i=\infty\} \log \pi_\infty (X_i,\hat{k}_i,\xi) \right).
\end{align*}
Using $\hat{\xi}$, the type-specific diff-in-diff estimators can be constructed as follows: \begin{align*}
\hat{\beta}_r(k) &= \sum_{e \leq T_1-1-r} \frac{\hat{\mu}(k,e)}{\sum_{e' \leq T_1-1-r} \hat{\mu}(k,e')} \cdot \widehat{ATT}_{e+r}(k,e)  
\end{align*}
where \begin{align*}
    \widehat{ATT}_t(k,e) &= \frac{\sum_{i=1}^n \left( Y_{it} -Y_{i,e-1} \right)\mathbf{1}\{\hat{k}_i=k,E_i=e\}}{\sum_{i=1}^n \mathbf{1}\{\hat{k}_i=k,E_i=e\}} \\
    &\hspace{5mm} - \frac{\sum_{i=1}^n \left( Y_{it} -Y_{i,e-1} \right) \mathbf{1}\{\hat{k}_i=k,E_i>t\} {\pi}_e (X_i,k,\hat{\xi})/{\pi}_{t+} (X_i,k,\hat{\xi})}{\sum_{i=1}^n \mathbf{1}\{\hat{k}_i=k, E_i>t\} {\pi}_e (X_i,k,\hat{\xi}) /{\pi}_{t+} (X_i,k,\hat{\xi})} \\
    \hat{\mu}(k,e) &= \frac{1}{n} \sum_{i=1}^n \mathbf{1}\{\hat{k}_i=k, E_i=e\} \\
    \pi_{e+} (x,k,\xi) &= \sum_{e' \geq e+1} \pi_{e} (x,k,\xi). 
\end{align*}

To discuss asymptotic properties of the type-specific diff-in-diff estimator, we adopt the following assumption:
\begin{assumption}
\hypertarget{A9}{} With some constant $M>0$, \begin{enumerate}[label=\textbf{\alph*.}]
\item (finite moments) For every $e$ and $t \geq -1$, $\E \left[ Y_{it}(e)^4 | k_i, X_{i} \right] \leq M$.

\item (type-specific parallel trend) For every $t,s \geq -1$ and $e$, \begin{align*}
\E \left[ Y_{it}(\infty) - Y_{is}(\infty) | k_i, X_i, E_i \right] &= \E \left[ Y_{it}(\infty) - Y_{is}(\infty) | k_i, X_i \right] \\
\E \left[ Y_{it}(e) - Y_{it}(\infty) | k_i, X_i, E_i\right] &= 0
\end{align*}
\item There exists some $\varepsilon^\pi > 0$ such that $\mu(k,e) > 0 \Rightarrow \Pr \left\lbrace \varepsilon^{\pi} \leq \inf_{w \in \Xi} \pi_e(X_i,k,w) \right\rbrace = 1$. 
\item Fix some $t,e$ and $k$ such that $\mu(k,e) > 0$ and $t \geq e$ and define a function $g : \Xi \to \mathbb{R}$ such that $$
g(w;X_i)= \frac{\pi_e(X_i,k,w)}{\pi_{t+}(X_i,k,w)}.
$$
There is a small neighborhood $B_\xi$ around $\xi$ with regard to $\| \cdot \|_2$ such that 

\textbf{i.} $g$ is almost surely twice continuously differentiable on $B_{\xi}$; \\
\textbf{ii.} $\frac{\partial}{\partial w} g(w)$ and $\frac{\partial^2}{\partial w \partial w^\intercal} g(w)$ are almost surely bounded by $M$ with regard to $\| \cdot \|_2$ on $B_{\xi}$. 

\end{enumerate}
\end{assumption}

\noindent Assumption \hyperlink{A9}{9-c} and \hyperlink{A9}{9-d} correspond to Assumption 7 of \citet{CS}. Note that Assumption \hyperlink{A9}{9-b} assumes a conditional parallel trend assumption for treatment periods, which is more relaxed the outcome model approach. That being said, the assignment model approach still assumes a linear parametric model for pretreatment periods; in this sense, the relaxation is intended for treatment effect heterogeneity and treatment timing assignment, not for untreated potential outcomes.

With Assumption \hyperlink{A9}{9}, we have the following corollary of Theorem \hyperlink{T2}{2}. 
\begin{corollary}
\hypertarget{C3}{} Assumptions \hyperlink{finite}{3}, \hyperlink{A7}{7}-\hyperlink{A9}{9} hold by replacing $X_{it}$ with $X_i$ and $m$ with $\pi_{e}$. There exists some $\nu^* >0$ such that $n/{T_0}^{\nu^*} \to 0$ as $n \to \infty$. Then, up to some permutation on $\{1,\cdots,K\}$, $$
\sqrt{n} \left( \hat{\xi} - \xi \right) \xrightarrow{d} \mathcal{N} \left( \mathbf{0}, \Sigma \right)
$$
with $\Sigma > 0$ from Assumption \hyperlink{A8}{8} as $n \to \infty$. In addition, the infeasible estimator $\tilde{\xi}$ admits an asymptotic linear approximation as follows: $$
\sqrt{n} \left( \tilde{\xi} - \xi \right) = \frac{1}{\sqrt{n}} \sum_{i=1}^n l^\pi(X_i,k_i,E_i) + o_p(1) 
$$
where $\E[l^\pi(X_i,k_i,E_i)]=0$ and $\E \left[ l^\pi(X_i,k_i,E_i) l^\pi(X_i,k_i,E_i)^\intercal \right] >0$. Then, up to some permutation on $\{1,\cdots,K\}$, $$
\sqrt{n} \left( \hat{\beta}_r(k) - \beta_r(k) \right) \xrightarrow{d} \mathcal{N} \left( 0, {\sigma_{\beta_r(k)}}^2 \right)
$$
with some ${\sigma_{\beta_r(k)}}^2 >0$, as $n \to \infty$, when $k$ and $r$ satisfy the corresponding overlap condition.
\end{corollary}

\section{Tables and figures}

\begin{table}[hp]
    \centering
    \small
    \renewcommand{\arraystretch}{1.5}
    \textsc{\caption{Simulation Results, $K=2$} \label{tab:sim1}} 
    \vspace{5mm}
    \begin{tabular}{c|ccccc}
        \hline 
        \multicolumn{6}{c}{Bias} \\
        \hline
        $(n,T_0)$ & DiD & SC & synthetic DiD & type-specific DiD & type-specific DiD \\
        \hline \hline
         $(50,10)$ & -0.540 & -0.107 & -0.183 & -0.008 & -0.017 \\
         $(50,20)$ & -0.594 & -0.129 & -0.110 & -0.027 & -0.027 \\
         $(50,30)$ & -0.571 & -0.140 & -0.090 & -0.035 & -0.035 \\
         \hline
         $(100,10)$ & -0.603 & -0.218 & -0.233 & -0.049 & -0.043 \\
         $(100,20)$ & -0.531 & -0.026 & -0.051 & 0.009 & 0.009 \\
         $(100,30)$ & -0.535 & 0.029 & -0.001 & 0.025 & 0.025 \\
         \hline
         Constant slope & - & - & - & NO & YES \\
         \hline 
    \end{tabular}
    \vspace{5mm} \\
    \begin{tabular}{c|ccccc}
        \hline 
        \multicolumn{6}{c}{MSE} \\
        \hline
        $(n,T_0)$ & DiD & SC & synthetic DiD & type-specific DiD & type-specific DiD \\
        \hline \hline
         $(50,10)$ & 0.696 & 0.620 & 0.435 & 0.370 & 0.367 \\
         $(50,20)$ & 0.754 & 0.642 & 0.371 & 0.342 & 0.342 \\
         $(50,30)$ & 0.753 & 0.662 & 0.380 & 0.363 & 0.363 \\
         \hline
         $(100,10)$ & 0.576 & 0.425 & 0.251 & 0.185 & 0.184 \\
         $(100,20)$ & 0.491 & 0.312 & 0.170 & 0.165 & 0.165 \\
         $(100,30)$ & 0.521 & 0.300 & 0.185 & 0.187 & 0.187 \\
         \hline
         Constant slope & - & - & - & NO & YES \\
         \hline 
    \end{tabular}
    \vspace{5mm} \\
    \begin{tabular}{c|rc|rc}
        \hline 
        \multicolumn{5}{c}{Classification success probability} \\
        \hline
        $(n,T_0)$ & \multicolumn{2}{c|}{$\leq 5$\% misclass.} & \multicolumn{2}{c}{No misclass.} \\
        \hline \hline
         $(50,10)$ & 0.984 & 1.000 & 0.748 & 0.904 \\
         $(50,20)$ & 1.000 & 1.000 & 1.000 & 1.000 \\
         $(50,30)$ & 1.000 & 1.000 & 1.000 & 1.000 \\
         \hline
         $(100,10)$ & 0.998 & 1.000 & 0.678 & 0.808 \\
         $(100,20)$ & 1.000 & 1.000 & 1.000 & 1.000 \\
         $(100,30)$ & 1.000 & 1.000 & 1.000 & 1.000 \\
         \hline
         Constant slope & NO & YES & NO & YES \\ 
         \hline 
    \end{tabular}
\end{table}

\begin{table}[hp]
    \centering
    \small
    \renewcommand{\arraystretch}{1.5}
    \textsc{\caption{Simulation Results, $K=3$} \label{tab:sim2}} 
    \vspace{5mm}
    \begin{tabular}{c|ccccc}
        \hline 
        \multicolumn{6}{c}{Bias} \\
        \hline
        $(n,T_0)$ & DiD & SC & synthetic DiD & type-specific DiD & type-specific DiD \\
        \hline \hline
         $(50,10)$ & -0.290 & -0.057 & -0.068 & -0.079 & -0.002 \\
         $(50,20)$ & -0.325 & -0.020 & -0.038 & -0.017 & -0.012 \\
         $(50,30)$ & -0.317 & -0.068 & -0.040 & -0.027 & -0.029 \\
         \hline
         $(100,10)$ & -0.304 & -0.046 & -0.077 & -0.084 & -0.024 \\
         $(100,20)$ & -0.321 & -0.050 & -0.039 & -0.018 & -0.017 \\
         $(100,30)$ & -0.329 & -0.005 & -0.031 & -0.029 & -0.029 \\
         \hline
         Constant slope & - & - & - & NO & YES \\
         \hline 
    \end{tabular}
    \vspace{5mm} \\
    \begin{tabular}{c|ccccc}
        \hline 
        \multicolumn{6}{c}{MSE} \\
        \hline
        $(n,T_0)$ & DiD & SC & synthetic DiD & type-specific DiD & type-specific DiD \\
        \hline \hline
         $(50,10)$ & 0.703 & 0.617 & 0.463 & 0.503 & 0.420 \\
         $(50,20)$ & 0.855 & 0.727 & 0.511 & 0.494 & 0.491 \\
         $(50,30)$ & 0.799 & 0.660 & 0.483 & 0.467 & 0.465 \\
         \hline
         $(100,10)$ & 0.443 & 0.434 & 0.261 & 0.274 & 0.232  \\
         $(100,20)$ & 0.428 & 0.330 & 0.215 & 0.211 & 0.211 \\
         $(100,30)$ & 0.445 & 0.316 & 0.224 & 0.224 & 0.224 \\
         \hline
         Constant slope & - & - & - & NO & YES \\
         \hline 
    \end{tabular}
    \vspace{5mm} \\
    \begin{tabular}{c|rc|rc}
        \hline 
        \multicolumn{5}{c}{Classification success probability} \\
        \hline
        $(n,T_0)$ & \multicolumn{2}{c|}{$\leq 5$\% misclass.} & \multicolumn{2}{c}{No misclass.} \\
        \hline \hline
         $(50,10)$ & 0.134 & 0.948 & 0.036 & 0.508 \\
         $(50,20)$ & 0.934 & 1.000 & 0.804 & 1.000 \\
         $(50,30)$ & 0.956 & 1.000 & 0.946 & 1.000 \\
         \hline
         $(100,10)$ & 0.316 & 0.988 & 0.028 & 0.250 \\
         $(100,20)$ & 0.996 & 1.000 & 0.970 & 1.000 \\
         $(100,30)$ & 1.000 & 1.000 & 1.000 & 1.000 \\
         \hline
         Constant slope & NO & YES & NO & YES \\ 
         \hline 
    \end{tabular}
\end{table}

\begin{table}[p]
    \centering
    \small
    \renewcommand{\arraystretch}{1.4}
    \textsc{\caption{Within-type Balancedness Test, $t=1988$} \label{tab:withintypebal}} 
    \vspace{3mm} 
    \begin{tabular}{c|ccc}
        \hline
        Type 1 & treated & never-treated & Diff \\
         \hline\hline
        $\mathbf{1}\{\text{central city}\}$ & 0.38 & 0.50 & -0.13 \\
        & (0.52) & (0.52) & (0.23) \\
        \% (white) & 59.12 & 63.26 & -4.15 \\
         & (17.83) & (20.59) & (8.37) \\
        \% (hispanic) & 7.26 & 4.21 & 3.05 \\
         & (12.81) & (7.04) & (4.91) \\
        \% (free/reduced-price lunch) & 39.36 & 35.65 & 3.71 \\
        & (10.07) & (17.49) & (5.87) \\
        \# (student) & 56604 & 62254 & -5650 \\
        & (38316) & (103167) & (30721) \\
         \hline
         N & 8 & 14 & - \\
         $p$-value & & & 0.827 \\
         \hline
    \end{tabular}
    \vspace{7mm} \\
    \begin{tabular}{c|ccc}
         Type 2 & treated & never-treated & Diff \\
         \hline\hline
        $\mathbf{1}\{\text{central city}\}$ & 0.62 & 0.73 & -0.12 \\
        & (0.51) & (0.46) & (0.18) \\
        \% (white) & 46.87 & 48.14 & -1.27 \\
         & (22.47) & (21.42) & (8.33) \\
        \% (hispanic) & 16.43 & 16.88 & 0.46 \\
         & (16.61) & (19.68) & (6.86) \\
        \% (free/reduced-price lunch) & 37.27 & 39.80 & -2.53 \\
        & (15.25) & (17.87) & (6.26) \\
        \# (student) & 74862 & 73790 & 1072 \\
        & (71857) & (154583) & (44612) \\
         \hline
         N & 13 & 15 & - \\
         $p$-value & & & 0.983 \\
         \hline
    \end{tabular}
    \vspace{3mm} \\
    \begin{minipage}{0.8\textwidth}
    \footnotesize
    The table reports means of the school district characteristics and their differences across treatment status within each type. The $p$-value is for the null hypothesis that the means of differences between treated units and never-treated units are all zeros. 
    \end{minipage}
\end{table}

\begin{table}[p]
    \centering
    \renewcommand{\arraystretch}{1.5}
    \textsc{\caption{Type-specific Descriptive Statistics, $t=1988$} \label{tab:bal}} 
    \vspace{3mm} 
    \begin{tabular}{c|cccc}
    \hline
         & Type 1 & Type 2 & Diff\\
         \hline\hline
        dissimilarity index & 29.94 & 37.63 & -7.69 \\
         & (13.39) & (18.53) & (4.52)  \\ 
        $\mathbf{1}\{\text{central city}\}$ & 0.45 & 0.68 & -0.22 \\
        & (0.51) & (0.48) & (0.14) \\
        \% (white) & 61.75 & 47.55 & 14.20 \\
         & (19.30) & (21.51) & (5.78) \\
        \% (hispanic) & 5.32 & 16.67 & -11.35 \\
         & (9.36) & (17.99) & (3.94) \\
        \% (free/reduced-price lunch) & 37.00 & 38.63 & -1.63 \\
        & (15.05) & (16.45) & (4.47) \\
        \# (student) & 60199 & 74288 & -14089 \\
        & (84178) & (121184) & (29096) \\
         \hline
         N & 22 & 28 & - \\
         $p$-value & & & 0.017 \\
         \hline
    \end{tabular}
    \vspace{3mm} \\
    \begin{minipage}{0.8\textwidth}
    \footnotesize
    The table reports the group means of the school district characteristics and their differences. The $p$-value is for the null hypothesis that the means of differences between Type 1 and Type 2 are all zeros. 
    \end{minipage}
\end{table}

\end{document}